\documentclass[prb,twocolumn,showpacs,&compress]{revtex4}
\usepackage{epsfig}
\usepackage{color}
\usepackage{amsmath, amsthm, amssymb}
\usepackage[latin9]{inputenc}
\usepackage{hyperref}
\usepackage{graphicx}
\usepackage{subfigure}

\newcommand{\bcen}{\begin{center}}
\newcommand{\ecen}{\end{center}}
\newcommand{\bec}{}
\newcommand{\btab}{\begin{tabular}}
\newcommand{\etab}{\end{tabular}}
\newcommand{\bdes}{\begin{description}}
\newcommand{\edes}{\end{description}}

\newcommand{\beq}{\begin{equation}}
\newcommand{\eeq}{\end{equation}}
\newcommand{\bea}{\begin{eqnarray}}
\newcommand{\eea}{\end{eqnarray}}

\newcommand{\half}{\frac{1}{2}}
\newcommand{\bary}{\begin{array}}
\newcommand{\eary}{\end{array}}
\newcommand{\benum}{\begin{enumerate}}
\newcommand{\eenum}{\end{enumerate}}
\newcommand{\exenum}{\begin{enumerate}[label=(\alph{*}),leftmargin=8mm]}
\newcommand{\bitem}{\begin{itemize}}
\newcommand{\eitem}{\end{itemize}}

\newcommand{\bk} { \mbox{\boldmath $k$}}

\newcommand{\bq} { \mbox{\boldmath $q$}}
\newcommand{\br} { \mbox{\boldmath $r$}}

\newcommand{\bK} { \mbox{\boldmath $K$}}

\newcommand{\eqn}[1] {Eqn.~(\ref{#1})}

\newcommand{\Fig}[1]{Fig.~\ref{#1}}

\def\tc{$T_c~$}

\def\c2{ CuO$_2~$}

\def\he4{${\rm {}^4He}$~}

\def\Gbar{\bar{G}}
\def\Abar{\bar{A}}

\newcommand{\tJ}{$t$-$J~$}

\newcommand{\citebyname}[1]{~\citeauthor{#1}~\cite{#1}}

\newcommand{\tk}{\tilde{\bf{k}}}
\begin{document}


\title{Role of the van Hove singularity in the Quantum Criticality of the Hubbard Model}


 \author{K.-S. Chen$^{*}$, S. Pathak$^{*}$, S.-X. Yang$^{*}$, S.-Q. Su$^{\dagger}$, D. Galanakis$^{**}$, K. Mikelsons$^{\ddagger}$, M. Jarrell$^{*}$ and J. Moreno$^{*}$}
 \affiliation{$^*$Department of Physics and Astronomy, Louisiana State University, Baton Rouge, LA 70803, USA\\$^{\dagger}$Oak Ridge National Laboratory, Oak Ridge, Tennessee 37831, USA\\$^{**}$Nanyang Technological University, Singapore 639798\\$^{\ddagger}$Department of Physics, Georgetown University, Washington DC, 20057, USA}

\date{\today}

\begin{abstract}
A quantum critical point is found in the phase diagram of the two-dimensional Hubbard model [Vidhyadhiraja \emph{et al.}, 
Phys.\ Rev.\ Lett.\ \textbf{102}, 206407 (2009)].  It is due to the vanishing of the critical temperature associated with a 
phase separation transition, and it separates the non-Fermi liquid region from the Fermi liquid.   Near the quantum 
critical point, the pairing is enhanced since the real part of the bare $d$-wave pairing susceptibility exhibits an 
algebraic divergence with decreasing temperature, replacing the logarithmic divergence found in a Fermi liquid 
[Yang \emph{et al.}, Phys.\ Rev.\ Lett.\ \textbf{106}, 047004 (2011)].  In this paper we explore the single-particle and 
transport properties near the quantum critical point using high quality estimates of the self energy obtained by 
{\em{direct}} analytic continuation of the self energy from Continuous-Time Quantum Monte Carlo.  We focus mainly 
on a van Hove singularity coming from the relatively flat dispersion that crosses the Fermi level near the quantum 
critical filling.  The flat part of the dispersion orthogonal to the antinodal direction remains pinned near the Fermi 
level for a range of doping that increases when we include a negative next-near-neighbor hopping $t'$ in the model. For 
comparison, we calculate the bare $d$-wave pairing susceptibility for non-interacting models with the usual two-dimensional 
tight binding dispersion and a hypothetical quartic dispersion. We find that neither model yields a van Hove singularity 
that completely describes the critical algebraic behavior of the bare $d$-wave pairing susceptibility found in the 
numerical data. The resistivity, thermal conductivity, thermopower, and the Wiedemann-Franz Law are examined in the Fermi 
liquid, marginal Fermi liquid, and pseudo-gap doping regions.  A negative next-near-neighbor hopping $t'$ increases 
the doping region with marginal Fermi liquid character.  Both $T$ and negative $t'$ are relevant variables for the quantum 
critical point, and both the transport and the displacement of the van Hove singularity with filling suggest that they are 
qualitatively similar in their effect.
\end{abstract}


\maketitle

\section{Introduction}
A plausible scenario for the high temperature superconductivity in cuprates is based upon the presence of a van Hove 
singularity corresponding to the saddle points in the single particle energy dispersion.~\cite{Labbe1987,Friedel1987, Newns1992, Markiewicz1997} 
These flat regions in the energy dispersion are directly observed in ARPES experiments on various cuprate 
compounds.\cite{Dessau1993, Gofron1994, Abrikosov1993, Campuzano1994, King1994} Recently, it was also observed in 
the tunneling spectra of Bi-2201.\cite{Piriou2011} The presence of saddle points in the energy dispersion is also argued 
to lead to a superconducting instability in other correlated systems, e.g. graphene\cite{McChesney2010}. If the Fermi 
level is doped to coincide with the van Hove singularity, then the superconducting transition temperature can be greatly 
enhanced. 

The van Hove scenario is also argued\cite{Pattnaik1992, Newns1994, Tsuei1990} to be responsible for the \emph{marginal 
Fermi liquid} behavior\cite{Varma1989} in which the lifetime broadening of the quasiparticles is of the order of its 
energy. Thus the van Hove scenario is argued to account for the linear-$T$ resistivity,\cite{Tsuei1990, Newns1994, Majumdar1996} 
$T$-independent thermopower,~\cite{Newns1994} anomalous isotope effect,~\cite{Tsuei1990} etc. 

There is numerical evidence for the presence of van Hove singularities in models of strongly correlated systems.  
The energy dispersion of one hole in an antiferromagnetic background has been considered in studies of the Hubbard 
model~\cite{Assaad1999, Imada1998} and the \tJ model.~\cite{Dagotto1994} These studies report the presence of extended 
saddle points. ~\citebyname{Assaad1999} found that the dispersion has a quartic dependence with momentum near the 
anti-nodal point $(\pi, 0)$.  

These examples of extended saddle points in various correlated superconducting systems, and their proximity to the 
Fermi level at the doping where the maximum transition temperature occurs, demonstrate that it is extremely important 
to understand the role played by these singularities. A plethora of scientific efforts have been devoted towards 
achieving this understanding.~\cite{Newns1991, Gopalan1992, Radtke94, Hlubina1995, Newns1995, Dzyaloshinskii1996, 
Djajaputra1997, Dias2000, Halboth2000, Kastrinakis2000, Irkhin2001, Irkhin2002, Sandeman2003, Katanin2004, 
Markiewicz2004, Kastrinakis2005, Shen2007, Shibauchi2008, Tamai2008, Zitko2009, Katanin2010, Schmitt2010} At the 
simplest level, the role of the van Hove singularity may be interpreted within the BCS formalism.  Here, the 
superconducting transition temperature, $T_c$, is determined by the condition $V \chi_0^\prime (\omega=0)=1$, 
where $\chi_0^\prime$ is the real part of the $ q=0$ bare pairing susceptibility, and $V$ is the strength of 
the pairing interaction.  
In a BCS superconductor, $\chi'_{0}(\omega=0)$ displays a logarithmic divergence as $T\to0$, yielding the BCS 
exponential form for \tc.  The van Hove singularity enhances the divergence of $\chi'_{0}(\omega=0)$,  yielding 
higher transition temperatures.

There is also strong evidence for a quantum critical point located beneath the superconducting 
dome in the cuprates, and in close proximity to the doping with the maximum \tc.~\cite{Sebastian2010,Taillefer2010} 
Above the quantum critical point, in a range of doping associated with marginal Fermi liquid behavior, the in-plane 
resistivity is known to vary linearly with $T$ over a wide range of temperatures.~\cite{Ando2004,Daou2009,Cooper2009,
Gurvitz1987,Wuyts1996,Uher1987,Schofield1999} In the Fermi liquid region the low temperature resistivity varies as 
$T^2$. The resistivity increases as the doping decreases from the Fermi liquid into the pseudogap region. 
Moreover, the thermal conductivity $\kappa$,~\cite{Sun2003,Williams1998} the thermopower 
$S$~\cite{Phillips2010,Trodahl1995} and the tunneling conductance $g$~\cite{Gurvitch1989} have been investigated near 
the quantum critical point of the cuprates.  $\kappa$ is 
observed to be nearly independent of temperature in the marginal Fermi liquid state~\cite{Muller1988} and 
depends on $1/T$ in the Fermi liquid region. This is consistent with the Wiedemann-Franz Law,\cite{Wiedemann1853} 
$\kappa \rho \propto T$. \citebyname{Phillips2010} suggested that the thermopower changes sign abruptly 
near the optimal doping in most of the cuprate materials, signaling a state with particle-hole symmetry. 
Also in the marginal Fermi liquid, the tunneling conductance $g(V)\sim g_0+g_1|V|$, where $g_0$ and $g_1$ 
weakly depend on $T$ and $V$.

\begin{figure}
\centerline{\includegraphics[width=3.2in]{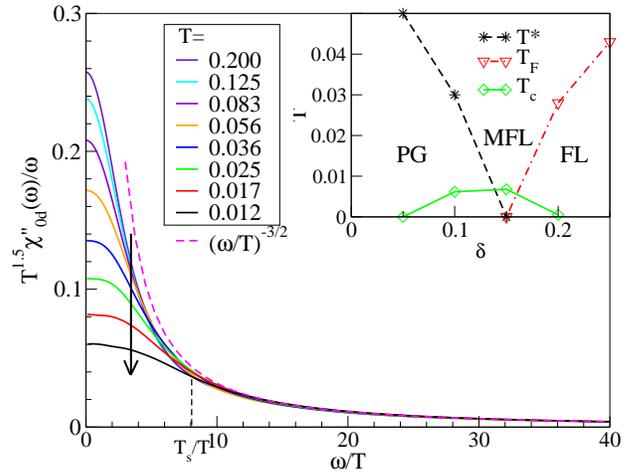}}
  \caption{ (color online) Frequency dependence of the imaginary part of the particle-particle $d$-wave susceptibility 
obtained from the Dynamical Cluster Approximation for various temperatures at the critical filling $n = 0.85$ when 
$U=6t$, $t'=0$ and $4t=1$.  The arrow denotes the direction of decreasing temperature. All the curves fall on top of each 
other for frequencies greater than $T_s/T \approx 4t/J$. The inset shows the phase diagram for the same parameters
($t$, $t'$ and $U$)
including the Fermi Liquid (FL), Marginal Fermi Liquid (MFL) and Pseudogap (PG) regions.  The lines indicate 
the $d$-wave transition temperature $T_c$ (determined by extrapolation of data from $N_c=8$, $12$ and $16$ site
clusters), the pseudogap temperature $T^*$, and the Fermi liquid temperature $T_F$.  (Taken from~\citebyname{Yang2010})  
The quasiparticle fraction on the Fermi surface vanishes at the quantum critical point where $T^*=T_F=0$, and remains 
zero in the pseudogap region.~\cite{Vidhyadhiraja2009}
 }  
\label{chipp_DCA}
\end{figure}

A recent study~\cite{Vidhyadhiraja2009} reported the presence of a quantum critical point in the two-dimensional Hubbard model, where 
the quasiparticle spectral weight becomes zero. This quantum critical point separates the non-Fermi liquid pseudogap from the Fermi liquid 
region, and is surrounded by a superconducting dome (c.f.\ the inset in Fig.~\ref{chipp_DCA}). At finite temperatures, 
the Fermi liquid and pseudogap regions are separated by the marginal Fermi liquid.  Interestingly, at the quantum critical point, 
the density of states is found to be nearly particle-hole symmetric at low frequencies with a sharp peak 
at $\omega = 0$. This filling is tantalizingly close to the optimal doping where the superconducting transition 
temperature $T_c$ attains its maximum. The proximity of the superconducting dome to the quantum critical point was recently investigated 
by~\citebyname{Yang2010}.  Unlike the BCS case, they found that the bare $d$-wave pairing susceptibility $\chi'_{0d}(\omega=0)$ 
diverges algebraically as $\displaystyle \frac{1}{\sqrt{T}}$ at the quantum critical point, thus leading to a strongly enhanced 
$T_c$. Using the Kramers-Kr\"onig relation between the real part and the imaginary part of the susceptibility, 
$\displaystyle \chi'_{0d}(T) = \frac{1}{\pi}\int \frac{\chi''_{0d}(\omega)}{\omega} d\omega$,  the algebraic 
divergence of  $\chi'_{0d}(T)$ was found to come from a scaling behavior of the imaginary part 
$\chi''_{0d}(\omega)$. When $T^{3/2} \chi''_{0d}(\omega)/\omega$ is plotted against $\omega/T$, the different 
temperature curves fall on top of each other when $\omega \ge T_s \equiv 4tT/J$, determining a scaling function $H(x)$ such that  
$T^{3/2} \chi''_{0d}(\omega)/\omega = H(\omega/T) \approx (\omega/T)^{-3/2}$ (see \Fig{chipp_DCA}). The contribution 
from $H$ to $\displaystyle \chi'(T) = \frac{T^{-3/2}}{\pi}\int H(\omega/T) d\omega \propto T^{-1/2}$ which will 
dominate at low $T$.  

Since this enhanced behavior is expressed in the bare pairing bubble, dressed by the self energy but with no
vertex corrections, this discussion naturally raises the question about the role played by the van Hove 
singularity in the quantum criticality and its possible connection to the superconducting $T_c$.
In this manuscript, we use the dynamical cluster quantum Monte Carlo method to explore the 
relationship between the quantum critical point and the van Hove singularity for high-temperature superconductivity 
in the Hubbard model.  We obtain high quality estimates of the real-frequency
single-particle self energy $\Sigma(\bK,\omega)$ at the cluster momenta $\bK$ by direct analytic continuation of the Matsubara frequency
self energy $\Sigma(\bK,i\omega_n)$ using the maximum entropy method.~\cite{Jarrell1996,Wang2009} 
This direct method avoids the artifacts on the self energy that come about by inverting the 
coarse-grained Dyson's equation.~\cite{Jarrell1996} 
In the model without next-near-neighbor hopping ($t'=0$), we find that, as we dope the system 
across the quantum critical filling, a flat region in the dispersion crosses the Fermi 
level, accompanied by a sharp nearly symmetric peak in the density of states which also passes through the Fermi level.
 We find that the resistivity follows a linear-$T$ dependence over a wide range of temperatures 
yet a narrow range of doping (see~\Fig{fig:resis}). We use these high quality estimates of the  self energy to calculate the bare 
pairing susceptibility, we again find the collapse of the data found in \Fig{chipp_DCA}. To understand the role 
played by the van Hove singularity in determining this critical behavior, we have calculated the pairing 
susceptibility in the $d$-channel for two non-interacting models at half filling - the standard quadratic dispersion and a 
hypothetical quartic dispersion.  While the quartic dispersion can yield the observed algebraic divergence of 
$\chi'_{0d}(\omega=0)$, neither dispersion produces the collapse of the data found in \Fig{chipp_DCA}, suggesting that a van 
Hove singularity alone does not capture this phenomena.  For negative $t'$, 
the resistivity follows a linear-$T$ behavior over a wider range of doping, and the 
sharp peak in the density of states and the flat region of the dispersion linger near the Fermi level for the same wider 
range of doping.  These results suggest that the doping region affected by quantum criticality at low temperature becomes 
larger when $t'<0$. We also show that the zero-frequency imaginary part of the self energy $\Sigma^{\prime\prime}(T,\omega=0)$,
the dominant contribution to the resistivity, has a wider range of linear-$T$ behavior for $t'<0$ than $t'=0$.
All this motivates us to speculate a phase diagram near the quantum critical point in the discussion section.

This paper has been organized as follows.  Section~\ref{sec:formalism} briefly outlines the model and methods 
used in this study. Results are presented in section~\ref{sec:results}. Single particle properties are discussed 
in section~\ref{subsec:sing-part}; the pairing susceptibility calculation in section~\ref{subsec:susc}, the effect of 
$t'$ on the dispersion in~\ref{subsec:eff_tpm}, and transport results in~\ref{subsec:transport}. The broader implications of our results are 
discussed in section~\ref{sec:Discussion} and the paper is concluded in section~\ref{sec:Conclusion}.  

\section{Formalism}
\label{sec:formalism}

In this work, we look for direct evidence of the van Hove singularity and marginal Fermi liquid behavior in the 
spectra, electronic dispersion, and transport properties of the two-dimensional Hubbard model  
\begin{equation}
H=\sum_{\bk\sigma}\epsilon_{\bk}^{0}c_{{\bk}\sigma}^{\dagger}
c_{{\bk}\sigma}^{\phantom{\dagger}}+U\sum_{i}n_{{i}\uparrow}n_{{i}\downarrow},
\label{eq:hubbard}
\end{equation}
where $c_{{\bk}\sigma}^{\dagger}(c_{{\bk}\sigma})$ is the creation (annihilation)
operator for electrons with wavevector ${\bk}$ and spin $\sigma$,  
$n_{i\sigma} =c_{i\sigma}^{\dagger}c_{i\sigma}$ is the number operator, and the bare dispersion is given by
\begin{equation}
\epsilon_{\bk}^{0}=-2t\left(\cos k_{x}+\cos k_{y}\right)-4t'\left(\cos k_{x}\cos k_{y}-1\right)\, ,
\label{StdDisp}
\end{equation}
with $t$ and $t'$ being the hopping amplitude between the nearest-neighbor and the next-near-neighbor sites, 
respectively, and $U$ is the on-site Coulomb repulsion.

We employ the Dynamical Cluster Approximation (DCA)~\cite{Hettler1998, Hettler2000} with a quantum Monte Carlo  
algorithm as the cluster solver. The DCA is a cluster mean-field theory that maps the original lattice  
onto a periodic cluster of size $N_c=L_c^2$ embedded in a self-consistently determined host.  This many-to-one 
map is accomplished by dividing the lattice Brillouin zone into cells centered at momenta $\bK$, and coarse 
graining the lattice Green's functions by summing over the momenta labeled with $\tk$ within each cell 
\begin{equation}
\Gbar(\bK,\omega) = \frac{N_c}{N}\sum_{\tk} G(\bK+\tk,\omega),
\label{eq:Gbar}
\end{equation}
where $\Gbar$ and $G$ are the coarse-grained and the lattice single-particle propagators, respectively.  
The coarse-grained Green's function defines the cluster problem.  Spatial correlations up to a range $L_c$ within the 
cluster are treated explicitly, while those at longer length scales are described at the mean-field level. However the 
correlations in time, essential for quantum criticality, are treated explicitly for all cluster sizes.  To solve the 
cluster problem, we use continuous-time quantum Monte Carlo~\cite{Rubtsov2005}, which has no Trotter error,~\cite{Blumer2007} 
and the Hirsch-Fye quantum Monte Carlo method~\cite{Hirsch1986,Jarrell2001} for the charge polarizability 
in \Fig{fig:charge}. We employ the maximum entropy method~\cite{Jarrell1996} to calculate the real-frequency spectra.

\subsection{Calculation of Single-Particle Spectra}
\label{Formalism_MEM}
In previous calculations of the single-particle spectra, we analytically 
continue the quantum Monte Carlo $\Gbar(\bK,\tau)$ to obtain $\Gbar(\bK,\omega)$, and then invert 
the coarse-graining Eq.~(\ref{eq:Gbar}) to obtain the self energy $\Sigma(\bK,\omega)$.
This last step can introduce spurious features in $\Sigma(\bK,\omega)$. As observed previously,~\cite{Wang2009} 
it is better to analytically continue the self energy directly. However, the self energy spectra does not share 
the normalization of $\int d\omega \Abar(\bK,\omega) = 1$, where $\displaystyle \Abar(\bK,\omega) = 
-\frac{1}{\pi} \Gbar^{\prime\prime}(\bK,\omega)$. This normalization is a desirable feature since it allows us to 
treat the spectrum as a normalized probability distribution.  Since the Hubbard self energy
$\Sigma(\bK,i\omega_n) = \Sigma_H + U^2 \chi_{\sigma,\sigma}/i\omega_n +\cdots$, where 
$\chi_{\sigma,\sigma} = 
\left\langle n_\sigma n_\sigma\right\rangle -\left\langle n_\sigma \right\rangle^2
= n_\sigma(1-n_\sigma)$ 
is the local polarizability of a single spin species $\sigma$, and 
$\displaystyle \Sigma(\bK,i\omega_n) - \Sigma_H = -\frac{1}{\pi}\int d\omega  \frac{\Sigma^{\prime\prime}(\bK,\omega)}{i\omega_n-\omega}$. 
It is easy to see that the integral of $\Sigma(\bK,i\omega_n) - \Sigma_H$ is
$U^2 \chi_{\sigma,\sigma}$.  Therefore we will analytically continue 
\begin{equation}
{\frac{\Sigma(\bK,i\omega_n) - \Sigma_H}{U^2 \chi_{\sigma,\sigma}} }
= 
\int d\omega {\frac{\sigma(\bK,\omega)}{i\omega_n - \omega}},
\end{equation}
where $\displaystyle \sigma(\bK,\omega) = -\frac{1}{\pi} \Sigma^{\prime\prime}(\bK,\omega)/U^2 \chi_{\sigma,\sigma}$,
$\int d\omega {\sigma(\bK,\omega)} = 1$, using $\chi_{\sigma,\sigma}$ calculated
in the Monte Carlo process. After that we obtain the lattice self energy $\Sigma(\bk,\omega)$ by interpolating the cluster 
self energy $\Sigma(\bK,\omega)$ to get the single-particle spectral function $A(\bk,\omega)$.

\subsection{$d$-wave Pairing Susceptibility}

We calculate the susceptibility in the $d$-wave channel to the pair field ${\cal V} = -f_db^{\dagger}_d + \mbox{h.c}$, 
for various models with a van Hove singularity at the Fermi level. Here 
$b^{\dagger}_d = \half\sum_i\left(b^{\dagger}_{i+\hat{x}} - b^{\dagger}_{i+\hat{y}}\right)$ is the singlet creation 
operator, where $b^{\dagger}_{i+\hat{\alpha}}$ creates a singlet at bond $i$-$(i+\hat{\alpha})$, $\alpha = x, y$, and 
$f_d$ is a complex constant. The non-interacting $d$-wave pairing susceptibility $\chi_{0d}$ can be computed by 
calculating the susceptibility bubble
\bea
\chi_{0d}(T) = T \sum_{\bk,i\omega_n} g_d^2(\bk) G^0(\bk,i\omega_n)G^0(-\bk,-i\omega_n),
\eea
where $g_d(\bk)$ is the $d$-wave form factor given by $g_d(\bk) = \cos k_x - \cos k_y$. $G^0(\bk,i\omega_n)$ is the 
non-interacting Green function given by
\bea
G^0(\bk,i\omega_n) = \frac{1}{i\omega_n - \epsilon_{\bk}^{0}},
\eea
with $\epsilon_{\bf k}^0$ the bare band dispersion in Eq.~(\ref{StdDisp}). $\chi_{0d}$ can be evaluated using
standard Matsubara summation which gives
\bea
\chi_{0d}(T) = \sum_{\bk} g_d^2(\bk) \left(\frac{1-2f_{\bk}}{2\epsilon^{0}_{\bk}}\right),
\label{BareSusc}
\eea 
where $f_{\bk}$ is the Fermi function. 

\subsection{Transport Coefficients}
To explain the anomalous transport properties of the marginal Fermi liquid, Varma \emph{et al.}~\cite{Varma1989,Varma2002} 
postulate that for a wide range of wavevectors $\bq$, excitations make a contribution to the absorptive spin and charge 
polarizabilities reflected by
\begin{equation}
\chi''(\bq,\omega) \propto \min \left( \left|\omega/T \right|,1 \right) {\rm{sign}}(\omega).
\label{eq:MFL_chi}
\end{equation}
Electrons scattering from these excitations acquire a self energy
\begin{equation}
\Sigma(\bk,\omega) \propto \omega\ln\left(x/\omega_c \right) - i\pi x/2,
\label{eq:MFL_Sigma}
\end{equation}
where $x=\max(|\omega|,T)$, and  $\omega_c$ is a cutoff.  The marginal Fermi liquid ansatz has several consequences on 
experimentally relevant quantities, including transport anomalies, such as the linear-$T$ electrical resistivity, the 
tunneling conductance 
$g(V)\sim g_0+g_1|V|$, the photoemission, the nuclear relaxation rate $T_1^{-1}\sim aT+b$, the optical conductivity 
$\sigma(\omega)$, the Raman scattering, and the superconductive pairing. For the specific heat $C_v(T)$ and thermal 
conductivity $\kappa(T)$, Varma argued that the normal state's electronic contribution is hard to extract from the 
experimental data due to the large phonon contribution. The electronic thermal conductivity for the marginal Fermi liquid
approximates to a constant because the Wiedemann-Franz law roughly holds.

To calculate the various Onsager transport coefficients we use the Kubo formula:~\cite{Kubo1959} 

\begin{equation}
L^{ij}_{\alpha \beta} = \pi \int d\omega \left(- {{df}\over{d\omega}} \right) 
\omega^{i+j-2}~{\cal D}_{\alpha \beta}(\omega),
\label{eq:Onsager}
\end{equation}
where $f$ is the Fermi function and 
\begin{equation}
{\cal D}_{\alpha \beta}(\omega)=\frac{1}{N} \sum_{\bk} v^{\alpha}(\bk)v^{\beta}(\bk) A(\bk,\omega)^2,
\label{eq:EffDOS}
\end{equation}
where $v^{\alpha}(\bk)$ is the $\alpha$-component of the electron group velocity and 
$A(\bk,\omega)$ is the single-particle spectral function.  The different transport coefficients 
are given by combinations of $L^{ij}$.  For example, in units where $e=1$
and the chemical potential $\mu=0$, the resistivity $\rho(T)=1/L^{11}$, 
the thermopower $S=-L^{12}/T L^{11}$, the thermal conductivity 
$\displaystyle \kappa=\frac{1}{T}\left(L^{22} - (L^{12})^2/L^{11}\right)$, 
and the Peltier coefficient $\Pi=L^{21}/L^{11}$.

We note that a simpler estimate exists for the thermopower $S$.  Here, we perform a Sommerfeld expansion of $L^{12}$ 
at the Fermi level and get an alternative form:
\begin{equation}
S=-\frac{\pi^2}{3}T\frac{\partial\log[{\cal D}_{\alpha \beta}(\omega)]}{\partial \omega}\big|_{\omega =0}.
\label{eq:thermopower_del}
\end{equation}
If the electron group velocity is a constant, and the square of the single-particle spectra is approximated by 
$\delta(\omega-\epsilon_{\bk})\tau_{\bk}$, where $\tau_{\bk}$ is the relaxation time, also assumed constant, the thermopower over temperature 
becomes just the derivative of the logarithm of the density of states at the Fermi level.~\cite{Stovneng1990}

\section{Results}
\label{sec:results}
\subsection{Single Particle Properties for $t' = 0$}
\label{subsec:sing-part}
\begin{figure}
 \centerline{\includegraphics*[width=3.2in]{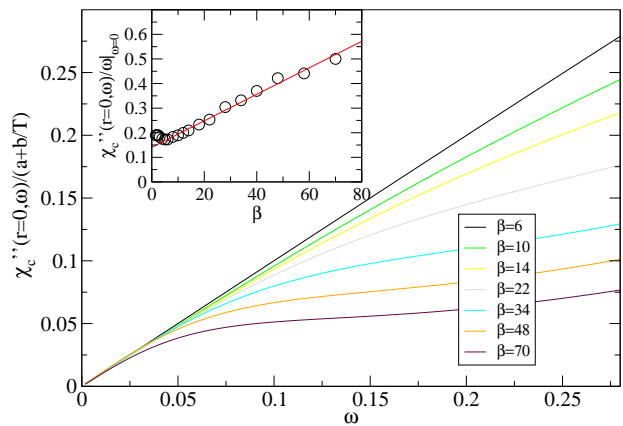}}
 \caption{(Color online) The local $\br=0$ imaginary part of the dynamical charge polarizability divided by the
initial slope at $\omega=0$, for $n=0.85$, $U=6t$, $4t=1$, $t'=0$ and $N_c=16$. It satisfies the marginal Fermi liquid 
form given by Eq.~(\ref{eq:MFL_chi}). Inset: the zero frequency slope of $\chi_c''(\br=0,\omega)$ is roughly 
linear in inverse temperature, as expected.  The line is a linear fit to the expression $a+b/T$.}
\label{fig:charge}
\end{figure}

We first explore the charge polarizability in the marginal Fermi liquid region at $n=0.85$. The imaginary component of 
the local charge polarizability $\chi_c''(\br=0,\omega)$ is plotted in Fig~\ref{fig:charge}.  The main plot shows 
$\chi_c''(\br=0,\omega)$ divided by its initial slope at zero frequency (determined in the inset), so that the curves 
coincide for low $\omega$.  At higher frequencies, the curves break from this linear rise at a frequency roughly proportional 
to the temperature. The inset shows that the zero frequency slope, $\left. \chi_c''(\br=0,\omega)/\omega\right|_{\omega=0}$, 
is roughly linear in inverse temperature up to $T\approx 0.2$ or roughly $2J=8 t^2/U$.  These features are consistent with 
the marginal Fermi liquid polarizability in Eq.~(\ref{eq:MFL_chi}). The spin polarization (not shown) does not display 
such an extended region of marginal Fermi liquid character.  This result is consistent with marginal Fermi liquid behavior 
due to the proximity of a quantum critical point associated with phase separation.

\begin{figure}
\includegraphics*[width=3.8in]{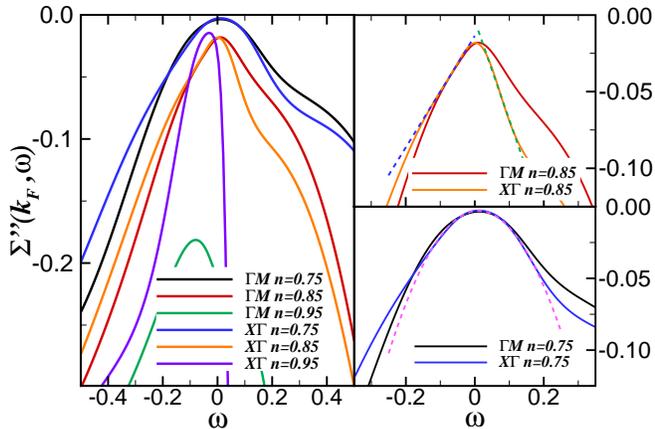}
\caption{(Color online) Frequency dependence of the imaginary part of the self energy at the Fermi momenta,
$\Sigma''(\bk_F,\omega)$, along $\Gamma$M and X$\Gamma$ for $U=6t$, $4t=1$, $t'=0$, $N_c=16$ and $\beta=58$. Right 
panels show a zoom of the low frequency region. Dashed lines fit the data linearly for $n=0.85$ and quadratically for $n=0.75$.} 
\label{fig:ImSigk_Fw}
\end{figure}

\begin{figure}
\includegraphics*[width=3.68in]{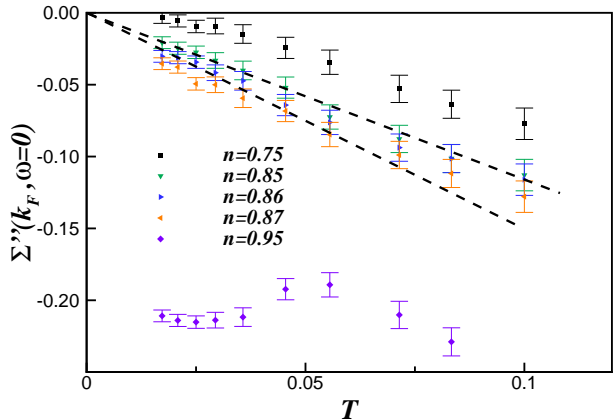}
\caption{(Color online) Temperature dependence of the imaginary part of the self energy $\Sigma''(\bk_F,\omega=0)$ 
at the Fermi energy and momenta, $U=6t$, $4t=1$, $t'=0$ 
and $N_c=16$. The linear dashed lines fit the self energies for $n=0.85$ and $0.86$ below $T=0.031$.} 
\label{fig:ImSigk_Fw0T}
\end{figure}

Fig.~\ref{fig:ImSigk_Fw} shows the frequency dependence of the imaginary self energy at the Fermi momenta along the 
anti-nodal (X$\Gamma$) and the nodal ($\Gamma$M) directions for three fillings: $n=0.75$ (Fermi liquid), $n=0.85$ 
(marginal Fermi liquid) and $n=0.95$ (pseudo-gap). For the self energy at $n=0.75$, the quadratic dashed line 
(bottom right panel)  provides a good fit to $\Sigma''(\bk_F,\omega)$ for small $\omega$, as expected from the Fermi 
liquid theory.~\cite{Hewson1993} The marginal Fermi liquid self energy has a form given by Eq.~(\ref{eq:MFL_Sigma}), 
which states that the imaginary self energy is proportional to the negative temperature for small frequency and to 
the negative $\omega$ when the temperature is low. The marginal Fermi liquid self energy in Fig.~\ref{fig:ImSigk_Fw} 
(upper right panel) shows a linear behavior, but interestingly with \emph{different} slopes for positive and 
negative $\omega$. This is not consistent with \eqn{eq:MFL_Sigma}, but this may be due to the presence of some 
short range order. I.e., \citebyname{Markiewicz2007} calculated the self-energy  due to the random-phase approximation 
(RPA) magnetic polarizability for a single band model with dispersion obtained by a fitting to the tight binding 
local-density approximation. They found that the self-energy has linear forms but with different slopes on positive and 
negative $\omega$ when the van Hove singularity is at the Fermi level and quadratic otherwise. 

Fig.~\ref{fig:ImSigk_Fw0T} shows the temperature dependence of the self energy when $\omega=0$ and $t'=0$. Again we find 
a result consistent with \eqn{eq:MFL_Sigma}, $\Sigma''(\bk_F,\omega\rightarrow 0) \propto -T$, around the marginal Fermi 
liquid filling for $n=0.85$ and $0.86$. The dashed lines are linear fits. The error bars are estimated by changing the 
random seeds in the Monte Carlo calculation. However, they do not reflect the systematic error that comes from the bias 
towards the default model, which in this case is the spectra from the next higher temperature.  This error is largest at 
low $T$ where the data is weak due to the minus sign problem.

\begin{figure}
\includegraphics*[width=3.8in]{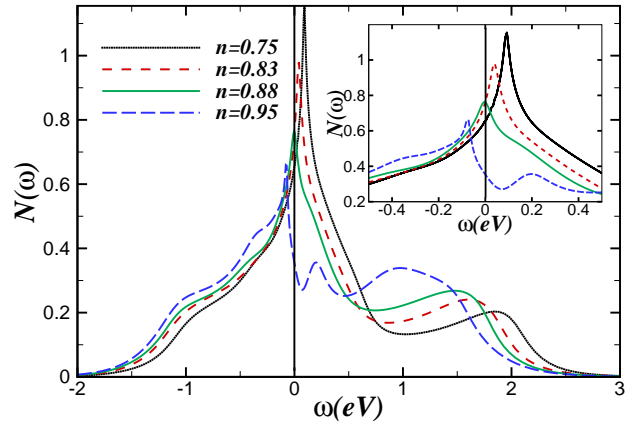}
\caption{(Color online) The single-particle density of states for $t'=0$, $U=6t$, $4t=1$, $N_c=16$ and $\beta=58$. 
The density of states shows low-energy particle-hole symmetry at the filling of $n=0.88$.} 
\label{fig:DOS}
\end{figure}

The self energy and dynamic charge polarizability near the critical filling $n_c=0.85$ are generally consistent 
with Varma's marginal Fermi liquid ansatz,  as are the results found previously for the kinetic and potential 
energies which vary with temperature like $T^2 \ln(T)$\cite{k_mikelsons_09b} and the vanishing wave 
function renormalization factor\cite{Vidhyadhiraja2009}. To understand the relationship of these results to the 
van Hove singularity, we will explore the density of states and the quasiparticle dispersion. 

The density of states for several fillings is shown in Fig.~\ref{fig:DOS}. Since we have highly 
enhanced the quality of the self energy by {\em{direct}} analytic continuation of $\Sigma(K,i\omega_n)$, 
the density of states in \Fig{fig:DOS} shows sharper features as compared to the results of \citebyname{Vidhyadhiraja2009}. 
As the doping decreases from the Fermi liquid to the pseudo-gap region, the peak of the density of states moves from 
positive to negative energy while its intensity is reduced. For $n=0.95$, a pseudogap begins to open and 
a peak to form at positive frequencies. The half-filled case ($n=1$, not shown) shows upper and lower 
Hubbard bands located at positive and negative frequencies, respectively. The density of states for filling $0.88$, 
close to the critical filling of $0.85$, shows low-frequency particle-hole symmetry.

\begin{figure*}
\centerline{\includegraphics[width=2.5in]{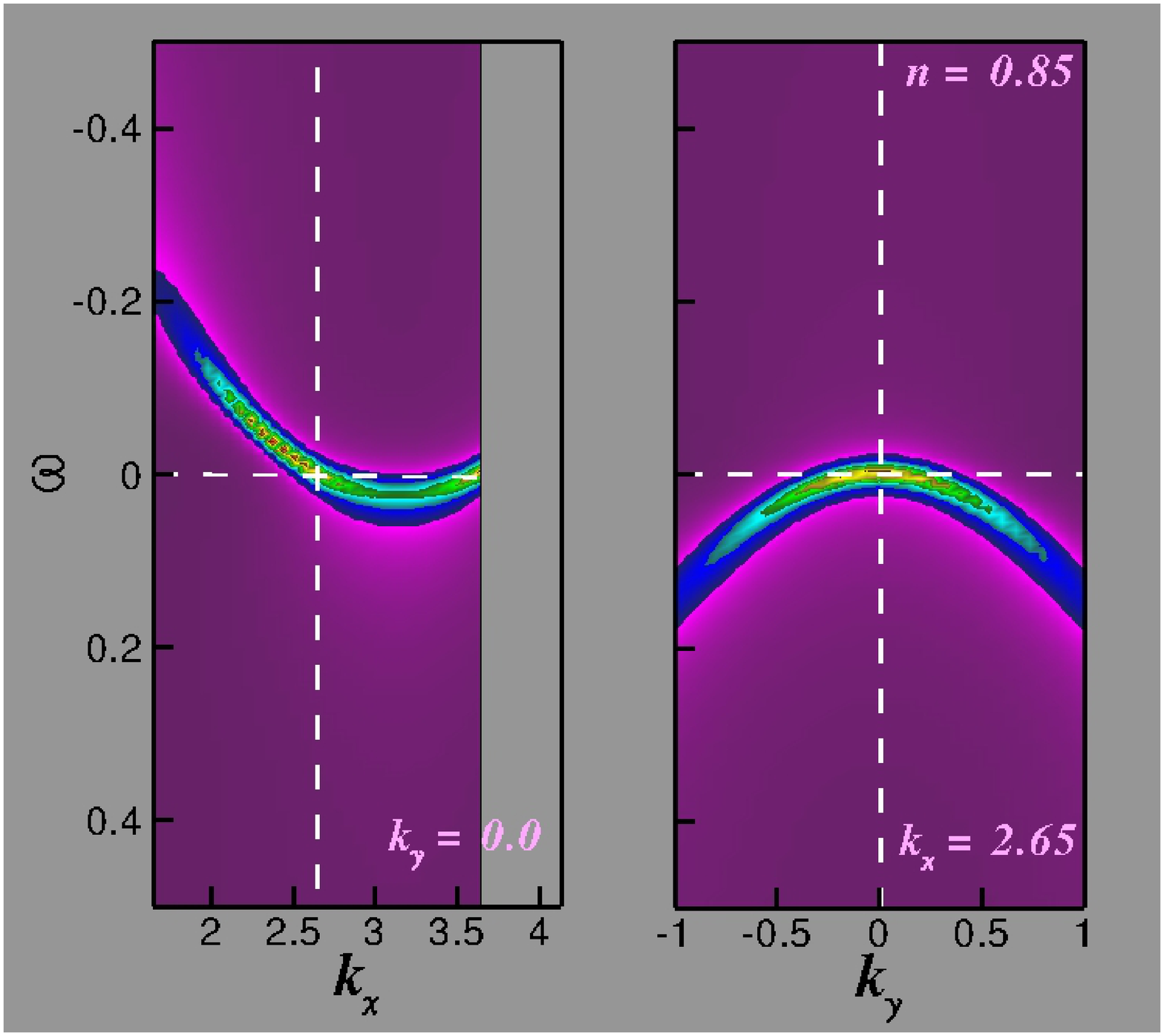}\includegraphics[width=2.5in]{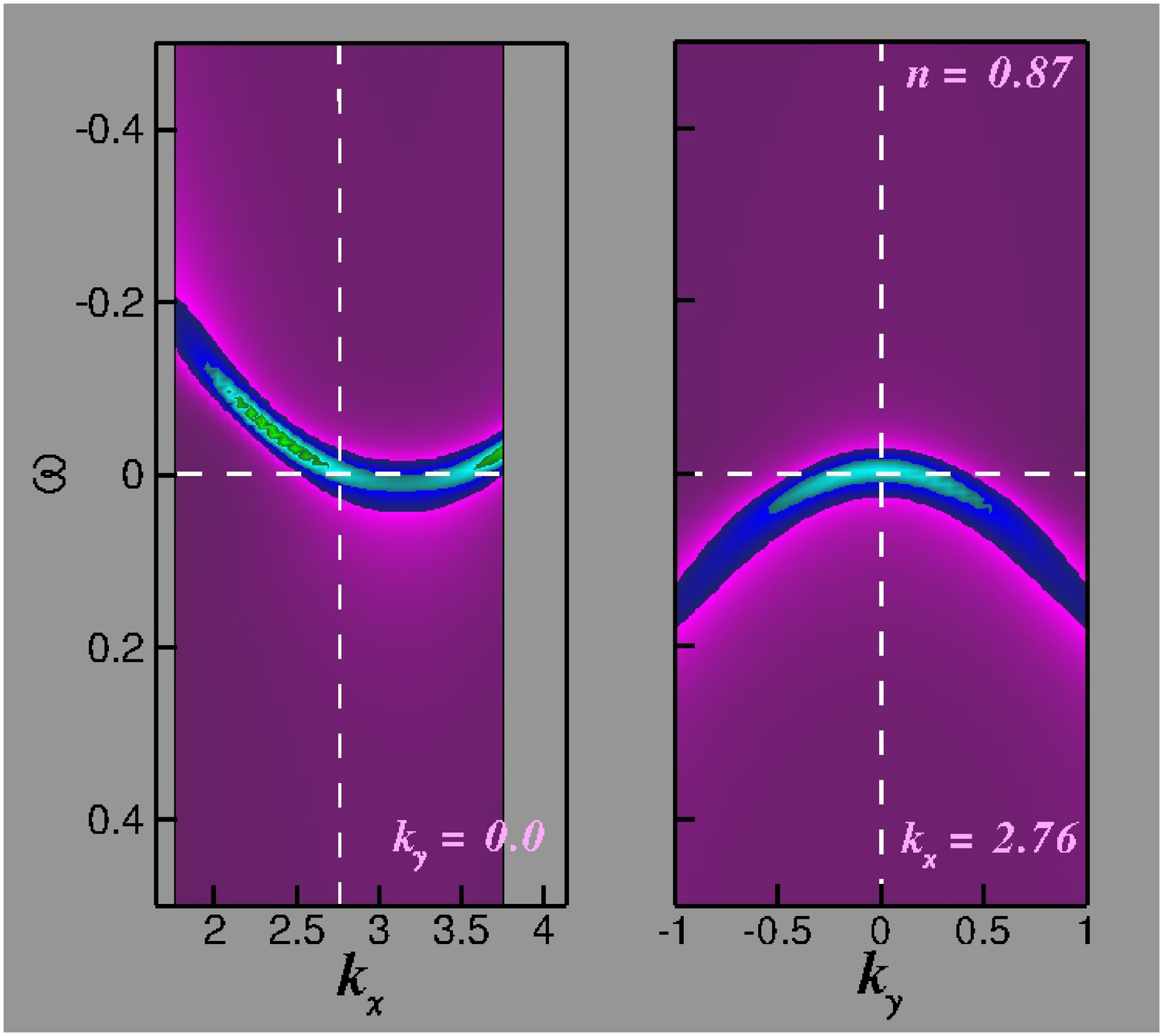}\includegraphics[width=0.65in]{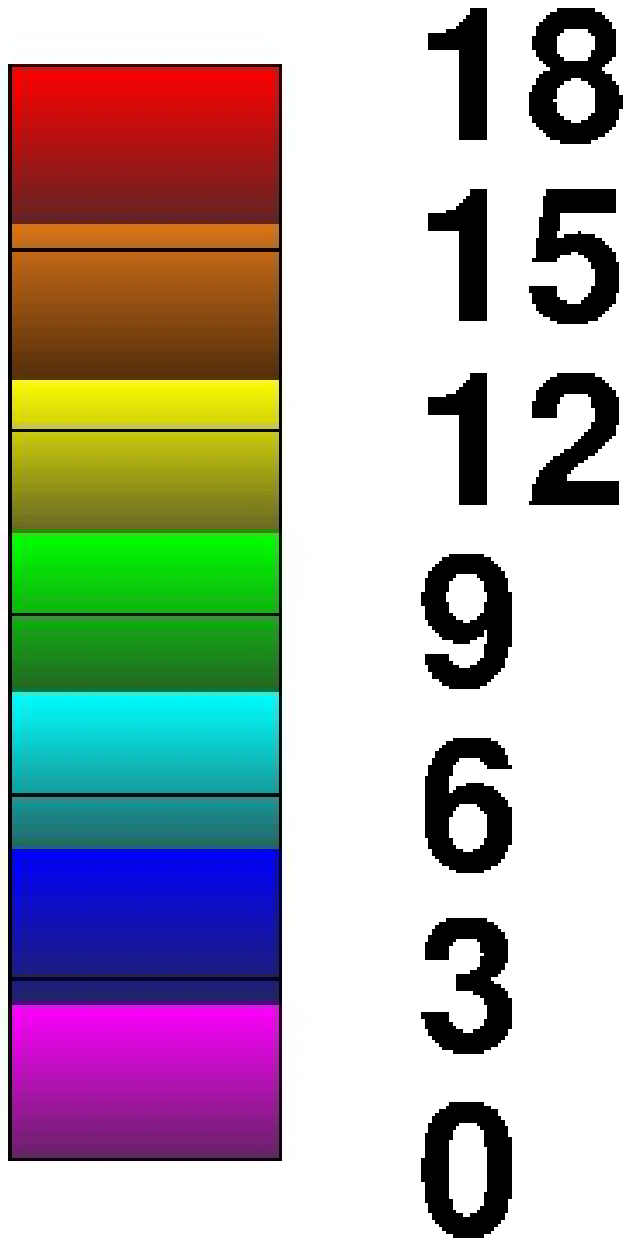}}
\centerline{\includegraphics[width=2.5in]{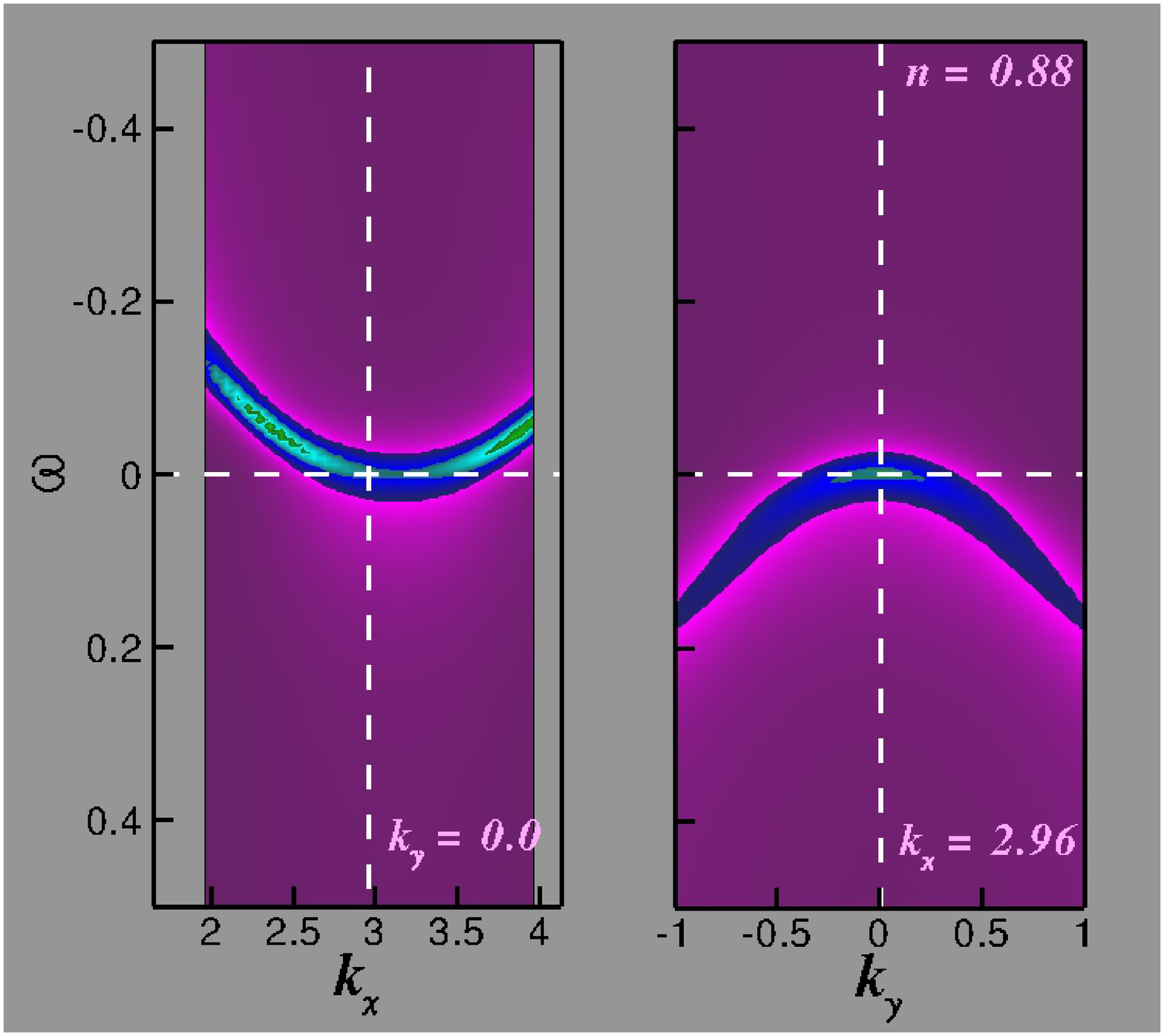}\includegraphics[width=2.5in]{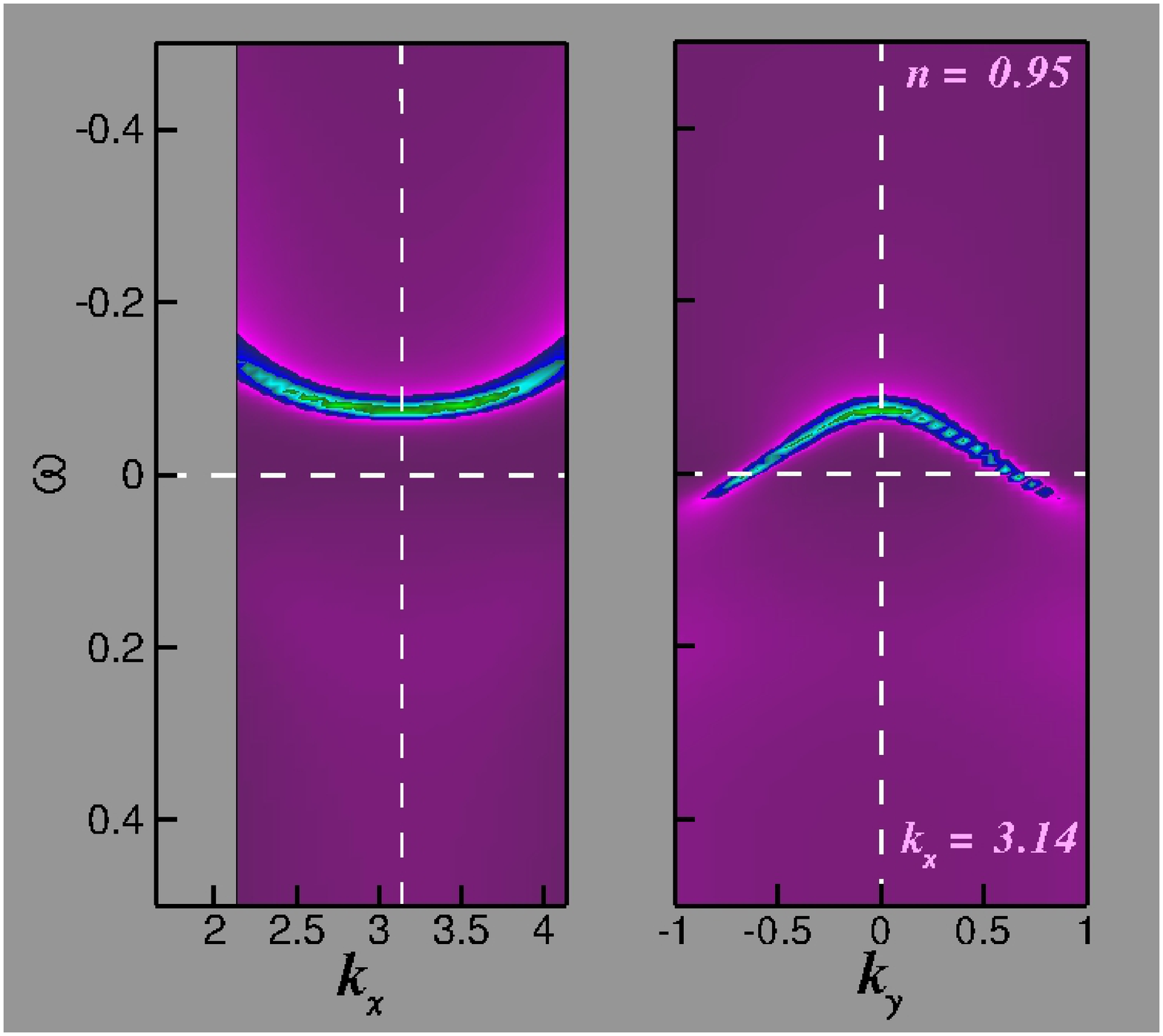}\hspace{0.65in}}
\caption{(color online) Energy dispersion obtained from the peaks of the spectral function $A(\bk,\omega)$ for various fillings 
around the Fermi vector $\bk_F$ along the anti-nodal direction for $t'=0$, $U=6t$, $4t=1$, $N_c=16$ and $\beta=58$.  By fixing $k_y=0$ 
we explore the dispersion along the $k_x$ direction, and for $k_x=k_{Fx}$ the dispersion along the $k_y$ direction is plotted. Notice 
that the energy axes are inverted so that positive energies are plotted down. The dispersion along the $k_y$ direction remains pinned 
near the Fermi level for a range of doping near the center of the superconducting dome (c.f.\ the inset of Fig.~\ref{chipp_DCA})
 }
\label{DispNearSadPt_blup}
\end{figure*}

Fig.~\ref{DispNearSadPt_blup} shows the dispersion obtained from the peaks of the spectral function $A(\bk,\omega)$ 
for four fillings: $n = 0.85$, $0.87$, $0.88$ and $0.95$, along the anti-nodal direction and around the Fermi vector 
$\bk_F$. In order to define $\bk_F$, we look for the maximum value of the zero frequency spectral function $A(\bk, \omega=0)$.  
In the Fermi liquid and marginal Fermi liquid regions, this definition is roughly equivalent
to the Luttinger surface defined where $G'(\bk, \omega=0)$ changes sign.  The two
definitions yield different results for the pseudogap region, especially when $t'<0$ which enhances the
pseudogap.   However, this difference is not large enough to qualitatively change our results or to change
any of our conclusions.   Therefore, for simplicity, we only show results using the first definition of $\bk_F$. For a particular 
filling, the left panel shows the dispersion along the $k_x$ direction ($k_y=0$), while the right panel shows the 
dispersion along $k_y$ ($k_x=k_{Fx}$). A common identifiable feature for all fillings is the presence of a 
flat region in the dispersion. This flat region is responsible for the van Hove singularity in the density of states. 
The van Hove singularity passes through the Fermi level at a filling of $n\approx 0.88$, which is near the quantum 
critical filling, $n\approx0.85$, where the quasi-particle weight $Z$ goes to zero.~\cite{Vidhyadhiraja2009} 
At this filling, the topology of the Fermi surface also changes from hole-like (closed around the point 
$\bk=(\pi,\pi)$) to electron-like (closed around $\bk=(0,0)$) with increasing filling (not shown) as seen in experiments.\cite{Kaminski2006}
The dispersion along the $k_y$ direction remains pinned near the Fermi level for a range of doping near 
the center of the superconducting dome, while the dispersion along the $k_x$ direction passes continuously
through the Fermi level.  This anisotropic motion of the flat dispersion is consistent with a van Hove
peak which moves continuously through the Fermi level as shown in Fig.~\ref{fig:DOS}, and would correspond 
to a flat region at the Fermi level which is most isotropic at the crossing and shrinks to narrow pencil-like 
regions for fillings above and below the crossing.

\begin{figure}
\centerline{
\includegraphics[width=3.8in]{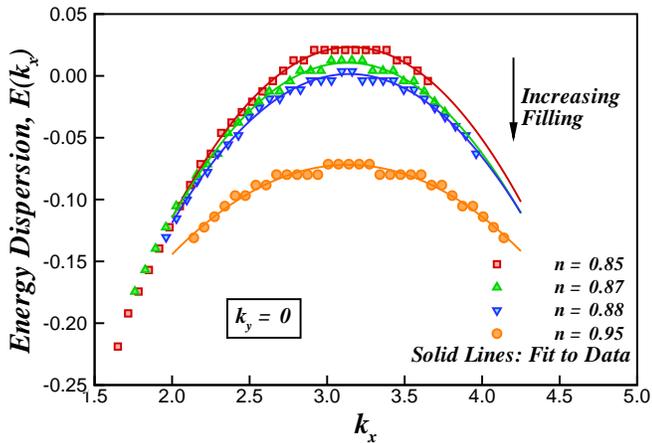}
}
  \caption{ (color online) Single particle dispersion around the Fermi energy taken along the anti-nodal direction
($k_y=0$). The data is from \Fig{DispNearSadPt_blup}.  The solid lines are fits to a quadratic dispersion.}
  \label{Disp_kx_various_fillings}
\end{figure}

The dispersion along the anti-nodal direction as a function of $k_x$ for various fillings is displayed in 
\Fig{Disp_kx_various_fillings}.  Interestingly, a quadratic form fits well to the data for all fillings. 
Next, we investigate if such a dispersion can capture the critical algebraic divergence of the pairing susceptibility.

\subsection{Pairing Susceptibility}
\label{subsec:susc}

\begin{figure}
\centerline{\includegraphics[width=3.5in]{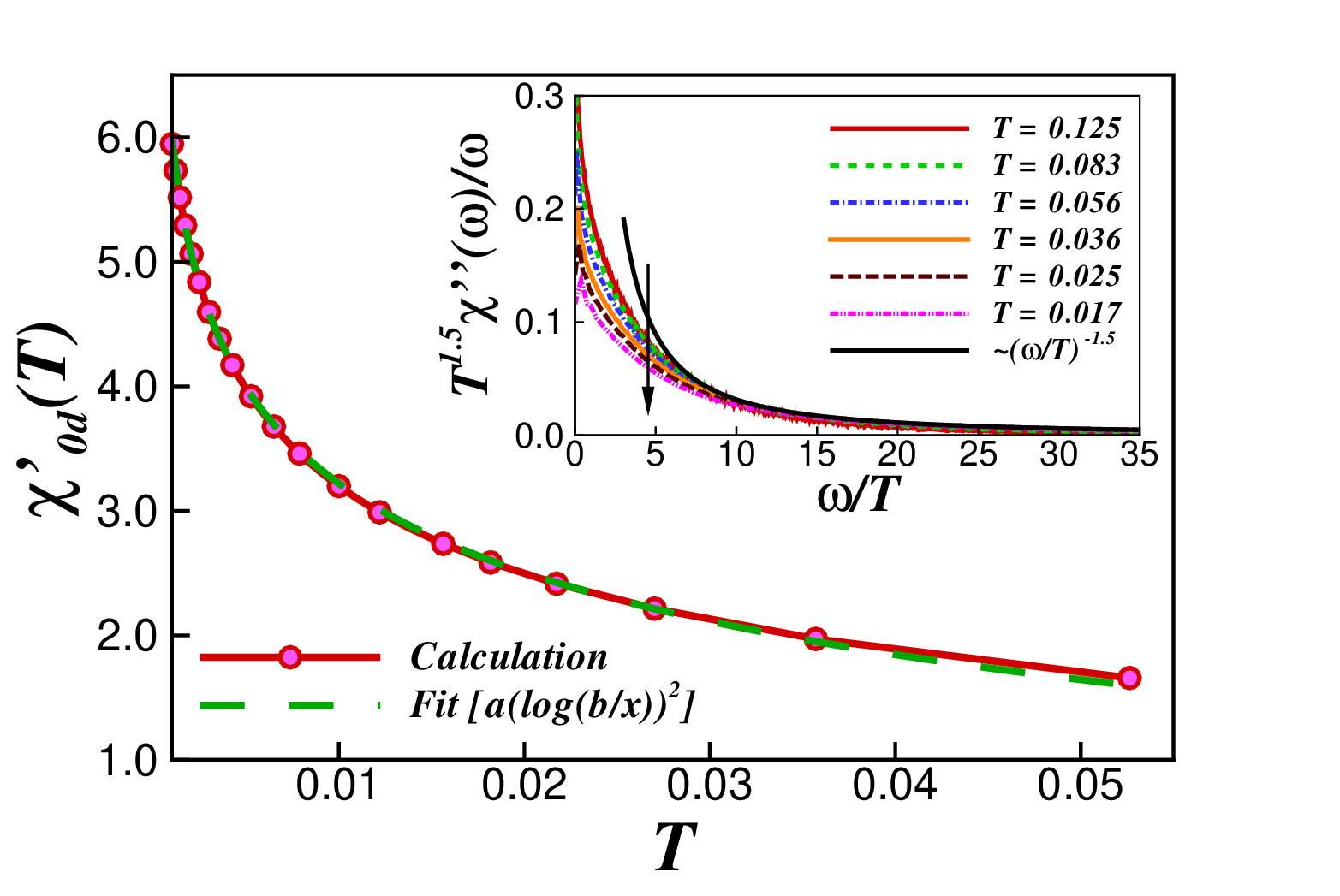}}
  \caption{ (color online) Temperature dependence of the real part of the particle-particle $d$-wave susceptibility at 
$\omega = 0$ for the two-dimensional tight binding dispersion at half filling. Note that $\chi'_{0d}$ diverges 
logarithmically as $T\to 0$. Inset: Frequency dependence of the imaginary part of the particle-particle $d$-wave 
susceptibility. Note that the curves corresponding to various temperatures do not scale at large frequency. The arrow 
denotes the direction of decreasing temperature.}
 \label{chi_quad}
\end{figure}

The density of states and the dispersion show clear evidence for a van Hove singularity which crosses the Fermi level near the
critical filling.  In order to see whether the van Hove singularity alone is sufficient to explain the enhanced
bare pairing bubble, we calculate the pairing susceptibility in the $d$-wave channel for two simple 
models having a van Hove singularity at the Fermi level. We begin with the tight binding model given by Eq.~(\ref{StdDisp})
at half filling and $t'=0$. The associated density of states has a logarithmic singularity at $\epsilon = 0$, 
$N(\epsilon) = \log |\epsilon|$. The temperature dependence of $\chi'_{0d}$ can be obtained by converting 
Eq.~(\ref{BareSusc}) to an integral over energy with a temperature $T$ cutoff. It results in a $-(\log T)^2$ behavior.
 This is confirmed by explicit calculation of the sum in Eq.~(\ref{BareSusc}) as illustrated in \Fig{chi_quad}. As shown 
in the inset, the imaginary part does not show scaling behavior as seen in \Fig{chipp_DCA}.

We also consider the next higher order model allowed by the symmetry of the square lattice, a hypothetical model 
with a quartic dispersion
\bea
\epsilon_{\bk} = -\frac{4}{\pi^4}\left((|k_x|-\pi)^4 - k_y^4 \right).
\label{eq:quartic}
\eea
Such an extended form has been observed in 
experiments~\cite{Gofron1994} and also confirmed by theoretical studies.~\cite{Imada1998} The low energy density of states 
for the quartic dispersion becomes $N(\epsilon)\sim 1/\sqrt{|\epsilon|}$.~\cite{Dias2000} Following a similar 
logic as we used for the tight binding dispersion, for a quartic dispersion, we get
$\displaystyle \chi'_{0d} \sim \frac{1}{\sqrt{T}}$.

\begin{figure}
\centerline{\includegraphics[width=3.5in]{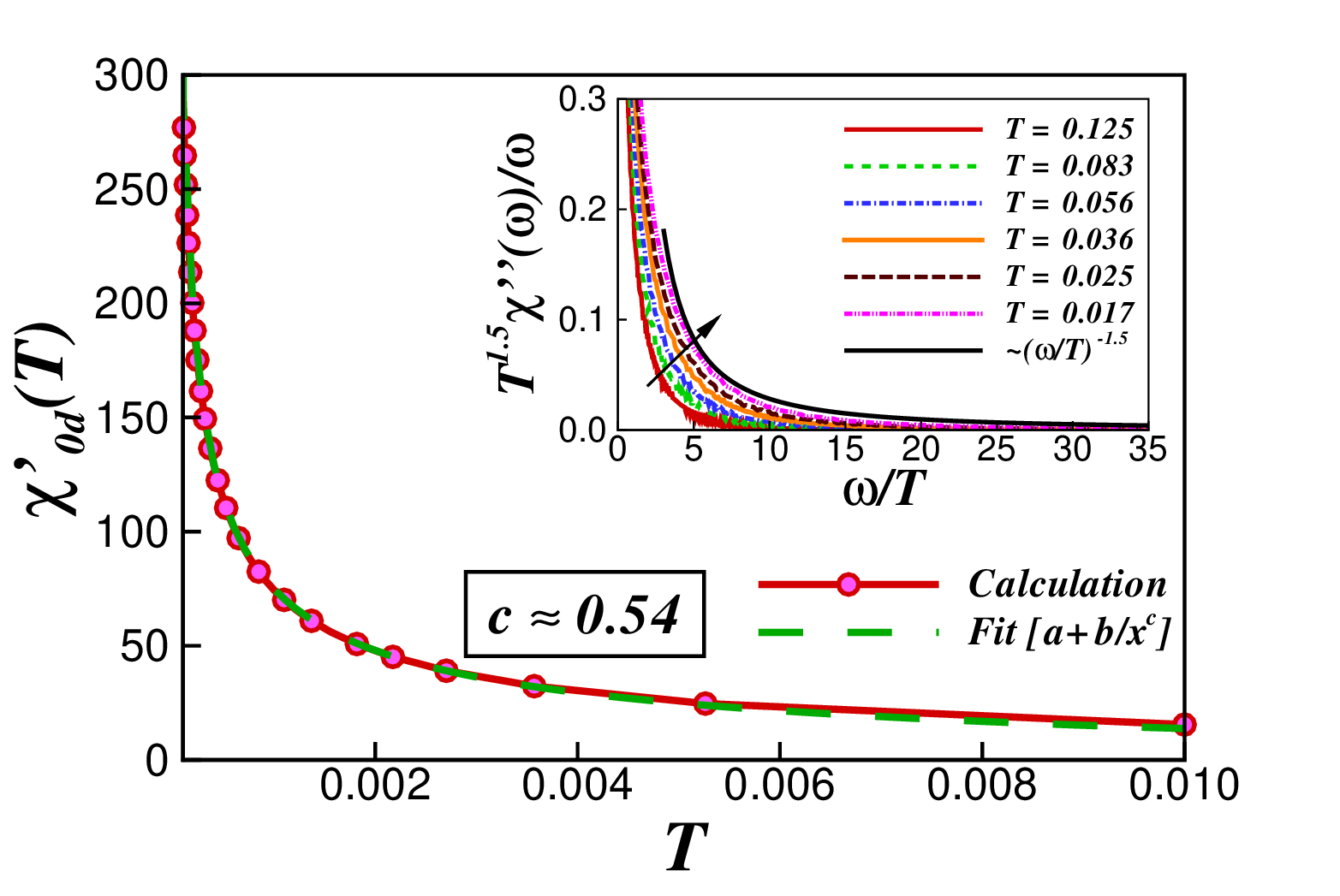}}
  \caption{ (color online) Temperature dependence of the real part of the particle-particle $d$-wave susceptibility 
at $\omega = 0$ for the quartic dispersion of Eq.~(\ref{eq:quartic}) at half filling. Note that $\chi'_{0d}$ diverges 
algebraically $\sim 1/\sqrt{T}$ as $T\to 0$. A fit to $a + b/x^c$ gives values of $a=-10.6$, 
$b=1.98$ and $c=0.54$.
Inset: Frequency dependence of the imaginary part of the pairing $d$-wave 
susceptibility. Note that the curves corresponding to various temperatures does not scale well at large frequency. The 
arrow denotes the direction of decreasing temperature.}
  \label{chi_quart}
\end{figure}

Results for the explicit calculation (Eq.~(\ref{BareSusc})) are shown in Fig.~\ref{chi_quart}, and are consistent with 
the analytical arguments above. Though the temperature dependence of the real part of the bare susceptibility is found 
to be algebraic for this quartic dispersion, the inset reveals that the imaginary pairing susceptibility 
does not exhibit the scaling found by \citebyname{Yang2010}. Thus, the simple non-interacting picture of the van Hove 
singularity at the Fermi level does not completely describe the true temperature and frequency dependence of the 
susceptibility.  

\subsection{Effect of negative $t'$}
\label{subsec:eff_tpm}

\begin{figure}
\includegraphics*[width=3.8in]{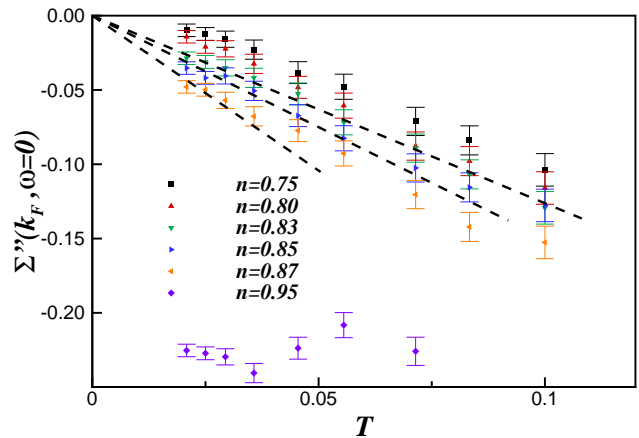}
\caption{(Color online) Temperature dependence of the imaginary part of the self energy $\Sigma''(\bk_F,\omega=0)$ at the 
Fermi energy and momenta,  $U=6t$, $N_c=16$, $4t=1$, and $t'/t=-0.1$. The self energy for filling between $n=0.83$ and 
$0.87$ shows a linear-$T$ behavior.} 
\label{fig:ImSigk_Fw0T_tp-0.1}
\end{figure}

\begin{figure*}
\centerline{\includegraphics[width=2.5in]{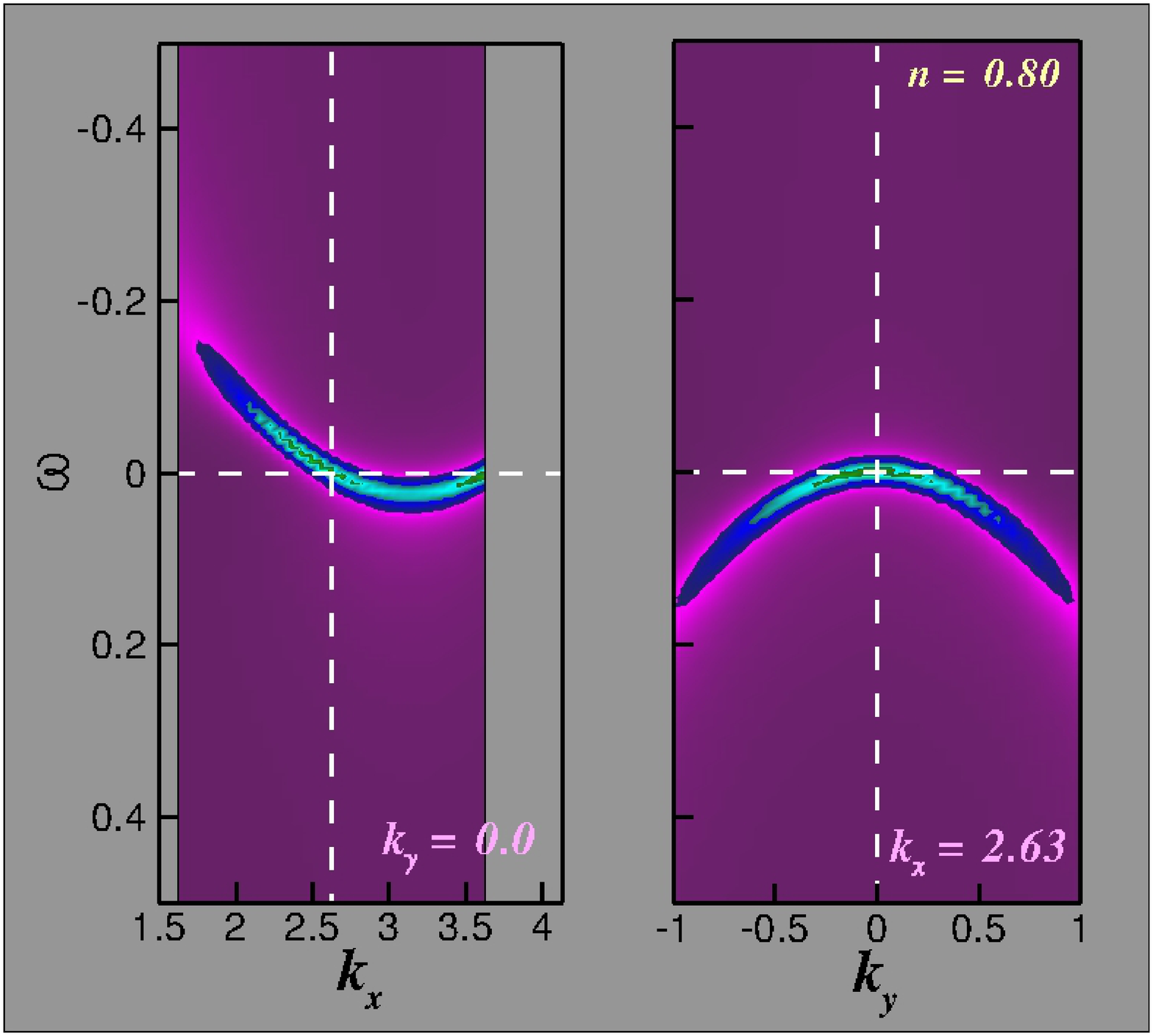}\includegraphics[width=2.5in]{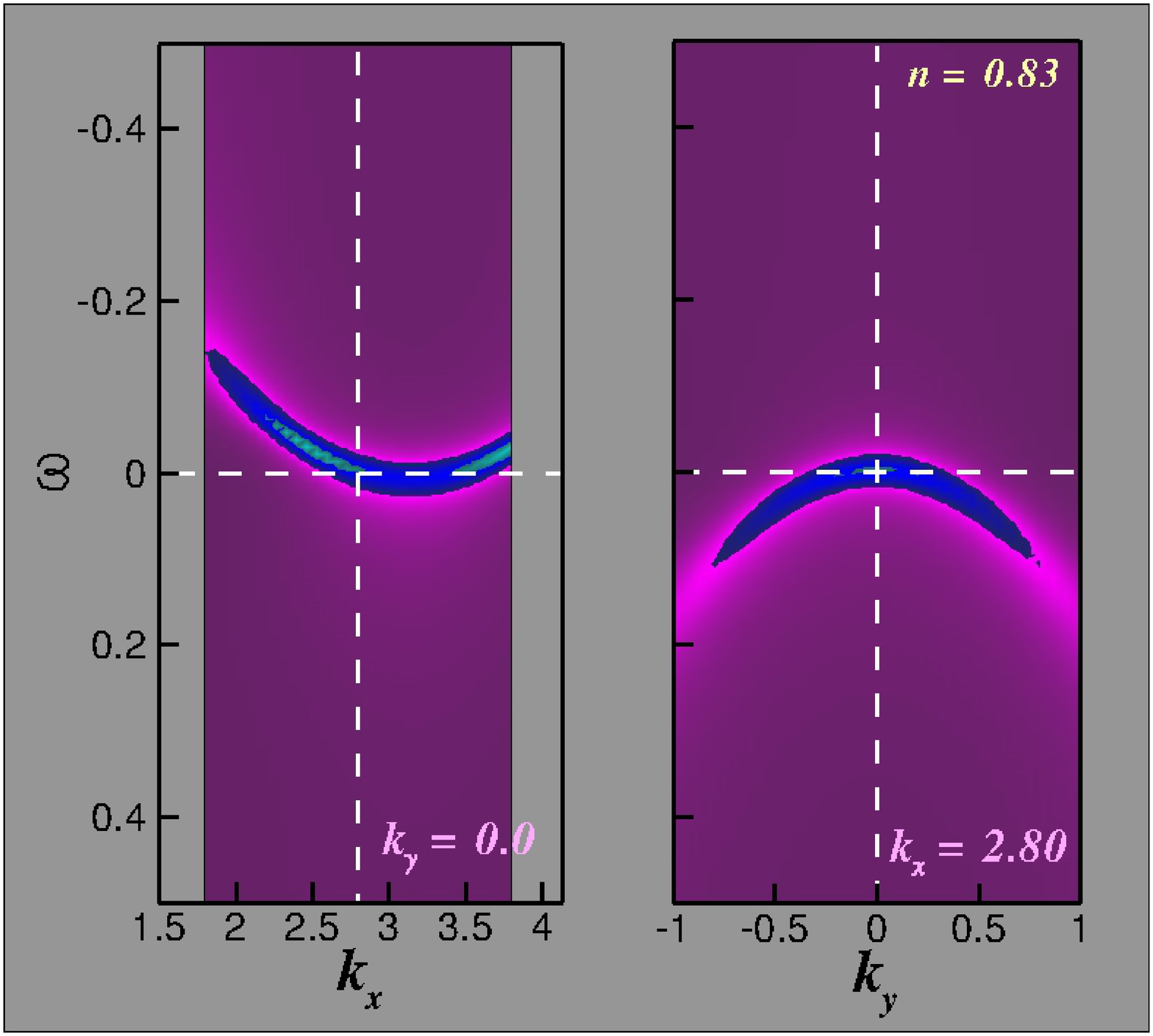}\includegraphics[width=0.65in]{legend.eps}}
\centerline{\includegraphics[width=2.5in]{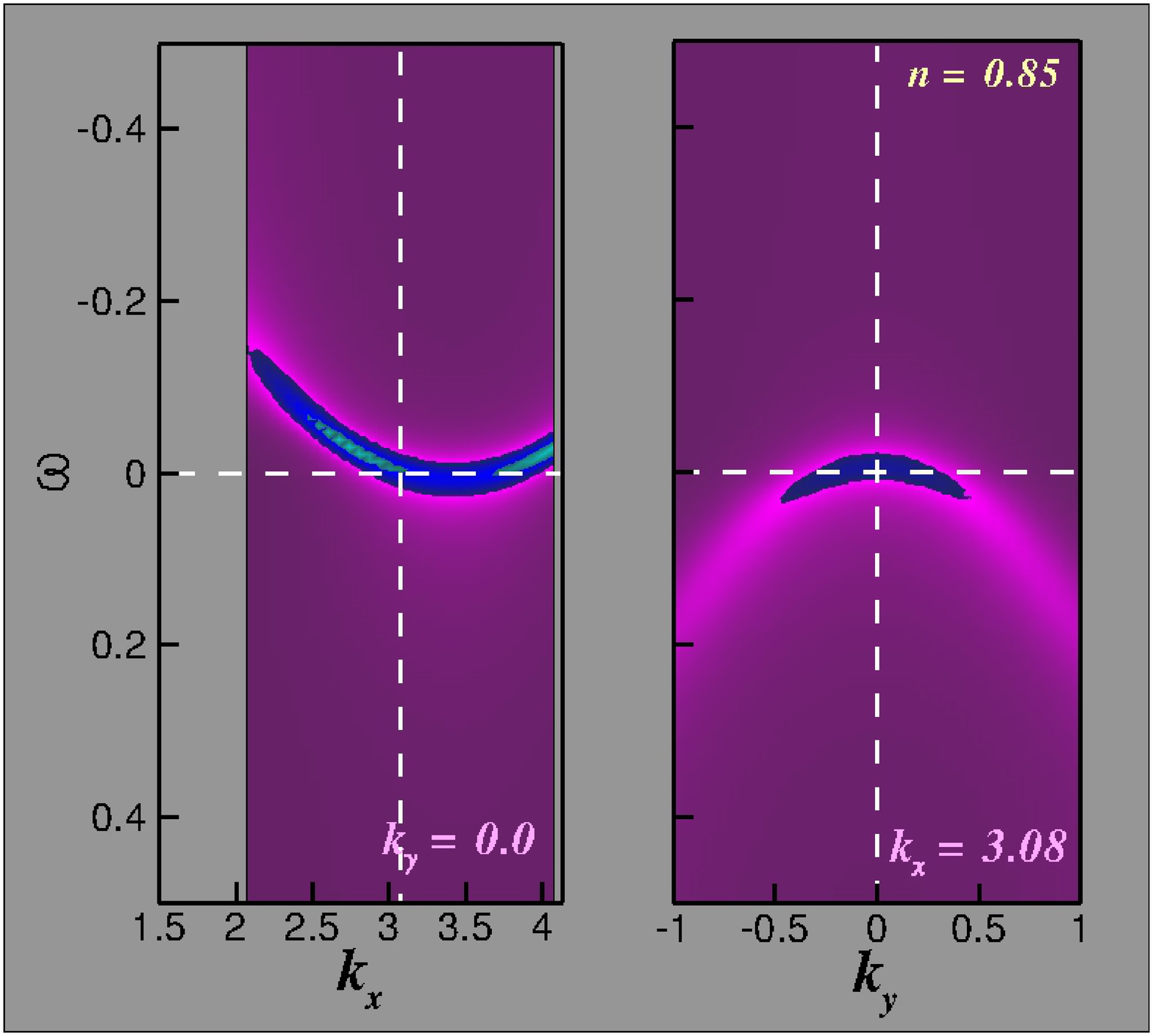}\includegraphics[width=2.5in]{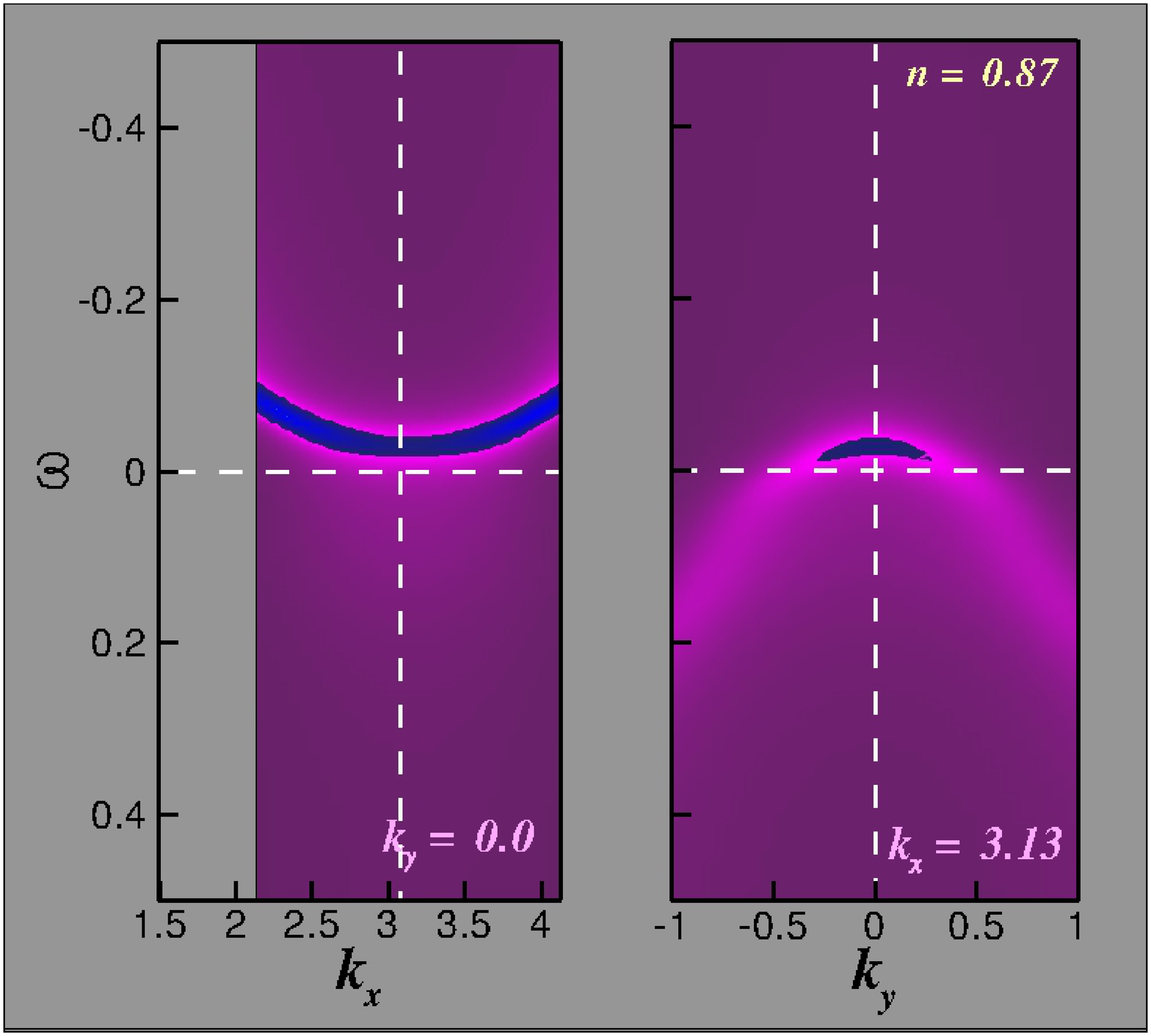}\hspace{0.65in}}
\caption{(color online) Energy dispersion obtained from the peaks of the spectral function $A(\bk,\omega)$ for various 
fillings around the Fermi vector $\bk_F$ along the anti-nodal direction for $t'/t=-0.1$, $U=6t$, $4t=1$, $N_c=16$ and $\beta=48$.
The dispersion along the $k_y$ direction remains pinned near the Fermi level for a broader range of dopings than 
found when $t'=0$.  Again, the dispersion along the $k_x$ direction moves continuously across the Fermi level.
}
\label{vHStpm}
\end{figure*}

\begin{figure}
\includegraphics*[width=3.8in]{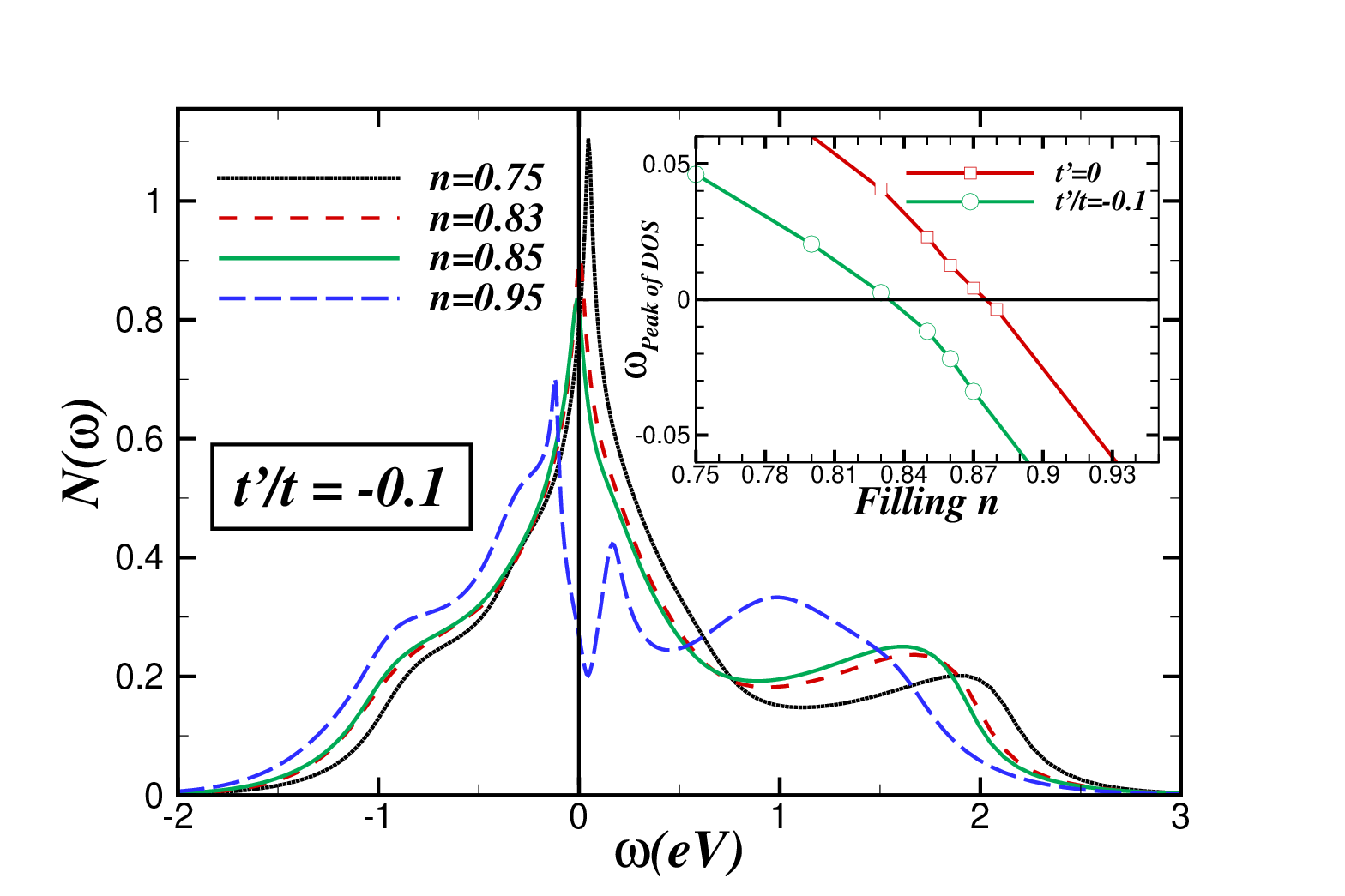}
\caption{(Color online) Single-particle density of states for $t'/t = -0.1$ when $U=6t$, $N_c=16$, $4t=1$, and $\beta=48$. 
Inset: Comparison of the filling dependence of the position of the peak of the density of states ($\omega_p$) for $t'=0$ 
and $t'/t=-0.1$. As the filling changes, $\omega_p$ for $t'=0$ crosses the Fermi level more quickly than for $t'/t=-0.1$.} 
\label{fig:DOS_tpm}
\end{figure}

The single-band Hamiltonian used to model the hole-doped cuprates generally includes a negative next-near-neighbor hopping 
$t'$. For $t'=-0.1 t$, the temperature dependence of the self energy at the Fermi momenta and energy is shown in 
Fig.~\ref{fig:ImSigk_Fw0T_tp-0.1}. We find that $\Sigma''(\bk_F,\omega=0)$ follows a linear behavior over a wider range of 
fillings, from $n=0.83$ to $0.87$. 

Fig.~\ref{vHStpm} demonstrates that the inclusion of a negative $t'$  also results in the pinning of the flat part of the
$k_y$-dispersion to the Fermi level.  However, now the pinning is observed for a larger range of fillings, roughly $0.80$ to 
$0.86$. Thus both measurements, the temperature dependence of the self energy and the pinning of the flat dispersion to 
the Fermi level, are consistent.  If we take the viewpoint that the quantum critical point and the pinning of the 
dispersion along $k_y$ to the Fermi level are concomitant aspects of quantum criticality, then a negative $t'$ leads 
to a larger range of 
quantum critical fillings. We will also see the signature of this behavior in various transport properties discussed in 
Section~\ref{subsec:transport}. 


\begin{figure}
\includegraphics*[width=3.2in]{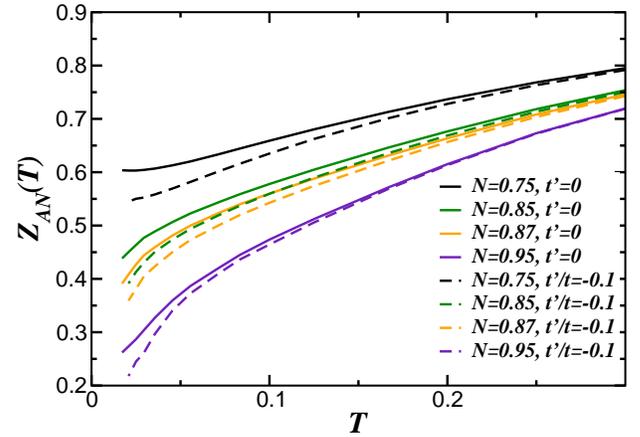}
\caption{(Color online) Temperature dependence of the Matsubara fraction along the anti-nodal direction, $Z_{AN}$, for 
various fillings for $t'=0$ and $t'/t = -0.1$. At the same filling and temperature, the Matsubara fraction 
decreases when $t'<0$.} 
\label{MatZ}
\end{figure}

\Fig{fig:DOS_tpm} shows the density of states for $t'/t =-0.1$ and various fillings as a function of $\omega$. For a given 
filling, the inset of Fig~\ref{fig:DOS_tpm} shows that the peak in the density of states is slightly shifted to smaller 
frequencies when compared with the peak in the density of states for $t'=0$. It displays particle-hole symmetry
roughly at $n=0.84$, not $n=0.88$ as for $t'=0$. Moreover, if we use $\Delta\omega_p/\Delta n$, where $\omega_p$ is 
the location of peak in the density of states and $n$ is the filling, to estimate the rate at which the peak crosses the Fermi level, 
we find that the peak of the density of states for $t'=0$ crosses the Fermi level more quickly than the peak  for negative $t'$. 
This can  be seen in the inset of Fig.~\ref{fig:DOS_tpm}, where the filling dependence of the peak location 
has a steeper slope for $t'=0$ than that for $t'/t=-0.1$ at the Fermi level. This confirms that 
negative $t'$ leads to a {\emph{wider}} range of fillings with a van Hove peak near the Fermi level.  The fact that
this range of fillings coincides with the region where marginal Fermi liquid behavior is seen in the self energy suggests 
that the van Hove singularity and quantum criticality are related.

Another interesting point to be noted here is that, when compared to the $t'=0$ result, the quasi-particle peaks become 
more incoherent for negative $t'$. This can be seen in the Matsubara quasiparticle weight along the antinodal momentum 
direction, $Z_{AN}$~\cite{Vidhyadhiraja2009}  displayed in Fig.~\ref{MatZ} as a function of temperature for different 
fillings. The quasiparticle fraction is consistently smaller for $t'/t = -0.1$ than for $t'= 0$ for all fillings.
This can also be seen through the increase of the blue color in the dispersion curves in Fig.~\ref{vHStpm} when compared
with Fig.~\ref{DispNearSadPt_blup}.

\subsection{Transport Properties}
\label{subsec:transport}

\begin{figure*}
\centerline{\includegraphics*[width=3.8in]{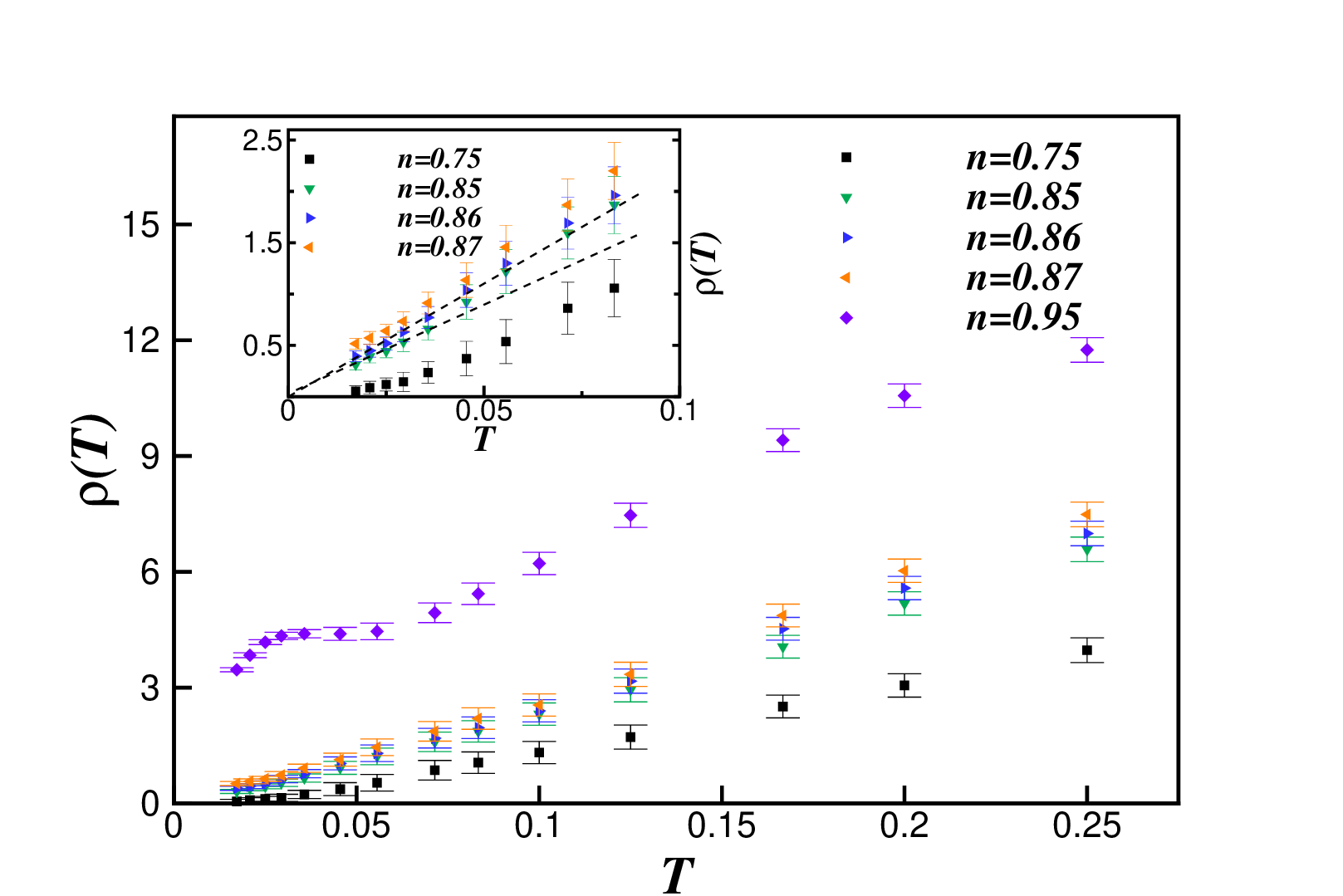}\includegraphics*[width=3.8in]{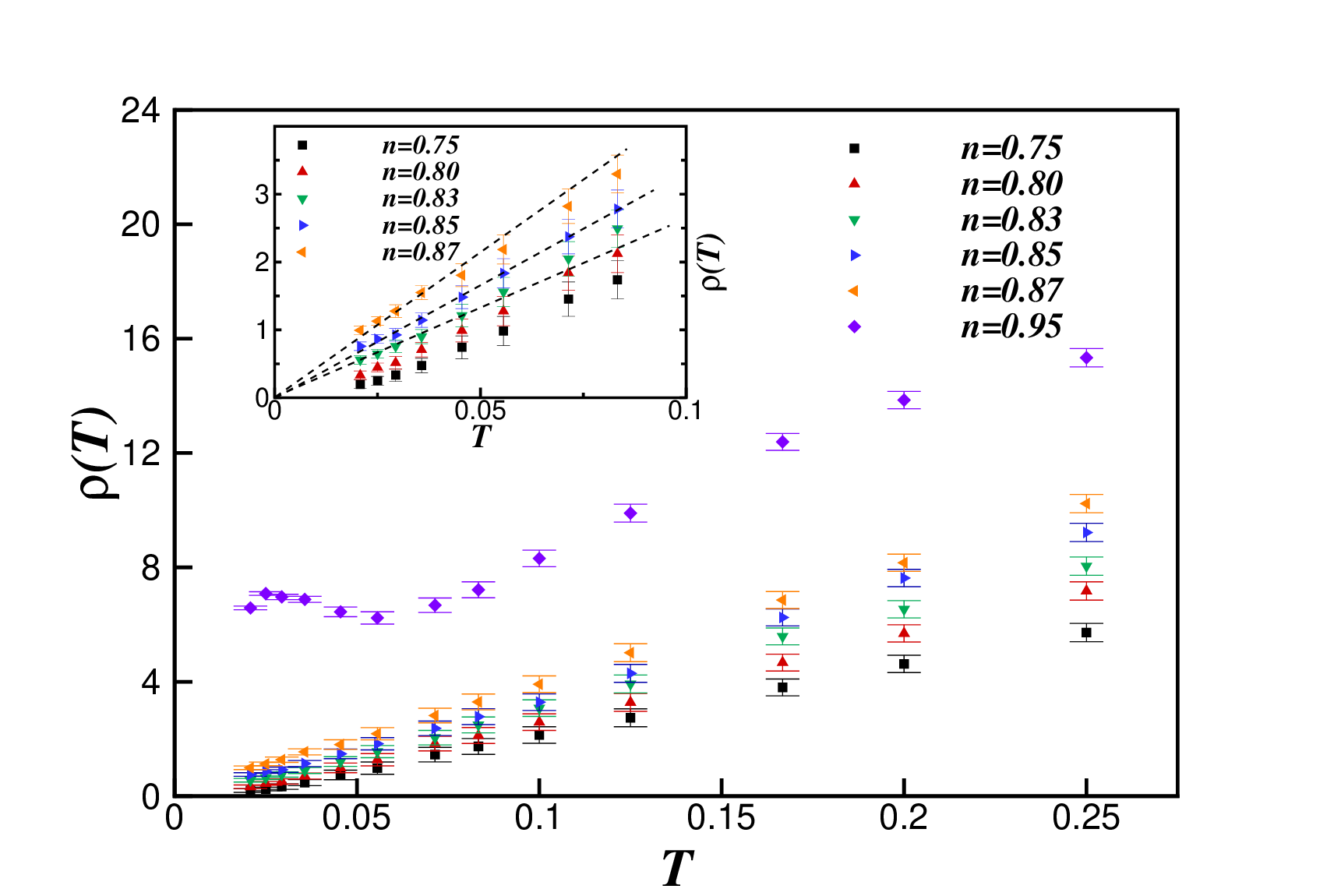}}
\caption{(Color online) Resistivity versus temperature for $t'=0$ (left) and $t'/t=-0.1$ (right) with $U=6t$, $N_c=16$, and 
$4t=1$ . The dashed lines in the insets are linear fits. For $t'=0$, the resistivity shows linear-$T$ behavior for $n=0.85$ 
and  $n=0.86$. For $t'/t=-0.1$, the resistivity shows linear-$T$ behavior from $n=0.83$ to $n=0.87$.} 
\label{fig:resis}
\end{figure*}

Matrix element effects\cite{Bansil1998,Bansil1999}, and the low precision of inverse photoemission can complicate
the direct measurement of the flat dispersion resulting in the van Hove singularity, making indirect probes like the Fermi 
surface topology\cite{Yoshida2001, Kaminski2006} and transport measurements more important.  The van Hove singularity and the 
quantum critical point will also impact the transport properties of the system. Using the Kubo formula under the 
relaxation time approximation in Eqs.~(\ref{eq:Onsager}) and (\ref{eq:EffDOS}), we obtain the resistivity, thermal 
conductivity, and thermopower in the Fermi liquid, marginal Fermi liquid, and pseudo-gap regions. 

\Fig{fig:resis} shows the resistivity as a function of temperature for $t'=0$, left panel, and $t'/t=-0.1$, right panel. 
Linear resistivity reveals evidence of the marginal Fermi liquid because the electronic cross section is proportional to 
$-\Sigma''(\bk_F,\omega=0)$ at low T, and, as seen in Fig. ~\ref{fig:ImSigk_Fw0T} and \ref{fig:ImSigk_Fw0T_tp-0.1}, this self 
energy is linear in $T$.  Again, for $t'=0$ a narrow range of fillings, from $n=0.85$ to $0.86$, displays a linear-$T$ 
resistivity at low $T$. While for $t'/t=-0.1$ a larger range of filling, $n=0.83$ to $0.87$, exhibits a linear temperature 
dependence. The linear resistivity in the 
marginal Fermi liquid region is consistent with experiments.~\cite{Ando2004,Daou2009,Cooper2009,Gurvitz1987} 
For $n=0.75$, both $t'=0$ and $t'/t=-0.1$ show Fermi liquid character, with a resistivity which goes to zero quadratically 
when $T$ approaches zero. The fact that doping region with marginal Fermi liquid character increases with negative
$t'$ has consequences for the phase diagram near the quantum critical point, which we will discuss in Section~\ref{sec:Discussion}. 

\begin{figure}
\centerline{\includegraphics*[width=3.8in]{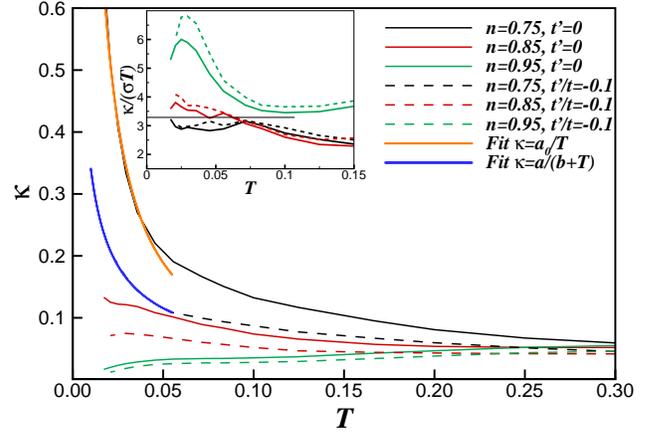}}
\caption{(Color online) Thermal conductivity versus temperature for $U=6t$, $N_c=16$, and $4t=1$.  Inset: Wiedemann Franz 
ratio for the same physical parameters. The horizontal solid line labels the constant $\pi^2/3$.} 
\label{fig:kappaT}
\end{figure}

\begin{figure}
\centerline{\includegraphics*[width=3.6in]{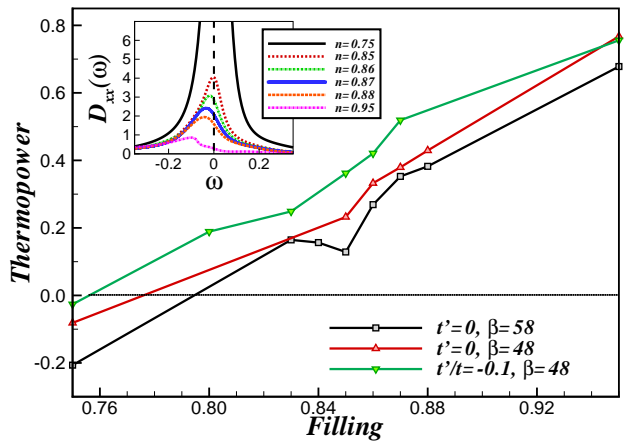}}
\caption{(Color online) Thermopower as a function of filling for $U=6t$, $N_c=16$, $4t=1$.
Lines are guides to the eyes.  Inset: Frequency dependence of ${\cal D}_{xx}(\omega)$ (c.f. Eq.~(\ref{eq:EffDOS}) for different fillings when 
$t'=0$, $\beta =58$. The slope of ${\cal D}_{xx}(\omega)$ at $\omega =0$ is proportional to the thermopower according to 
Eq.~(\ref{eq:thermopower_del}).} 
\label{fig:Thermopowerfilling}
\end{figure}

According to the Wiedemann-Franz Law, $\kappa/(\sigma T)=\pi^2/3$ ($k_{B}=e=1$), the thermal conductivity of a Fermi 
liquid is inversely proportional to $T$.~\cite{Wiedemann1853} \Fig{fig:kappaT} shows that $\kappa\propto 1/T$ for $n=0.75$ 
when $t'/t=0$ and $-0.1$, , but weakly depends on $T$ for the marginal Fermi liquid and the pseudo-gap regions.
The inset shows that, for $n=0.75$, the Wiedemann-Franz ratio $\kappa/(\sigma T)$ approaches a constant 
which is less than $\pi^2/3$ when $T\le 0.08$. ~\citebyname{Dahm1994} investigated the two-dimensional Hubbard model for 
$n\simeq 0.9$ and also found a smaller Wiedemann-Franz ratio.  However, we find that the Wiedemann-Franz ratio is larger 
than $\pi^2/3$ for the marginal Fermi liquid ($n=0.85$) and pseudo-gap ($n=0.95$) regions. We also see that the thermal 
conductivity becomes very small as $T\to 0$ for $n=0.95$ and saturates to a constant for $n=0.85$. So, when studying 
$\kappa$,  the marginal Fermi liquid seems to separate the Fermi liquid from the  pseudo-gap region. The dashed curves in  
Fig.~\ref{fig:kappaT} for $t'/t=-0.1$ data are always below the solid curves for $t'=0$ when plotting $\kappa$. However, 
the $t'/t=-0.1$ data is above the $t'=0$ results when we focus on  the ratio $\kappa/\sigma T$. This implies that negative 
$t'$ reduces the electrical conductivity more than the thermal conductivity.  


\citebyname{Phillips2010} argue that the thermopower changes sign near the quantum critical point, and that this is related 
with the development of a state with particle-hole symmetry. Fig.~\ref{fig:Thermopowerfilling} shows the thermopower $S$ 
as a function of filling. For $t'=0$ and $\beta=58$, the filling at which $S$ changes sign is roughly $0.80$. We expect 
that the zero-crossing of the thermopower will approach the critical filling of $0.85$ for decreasing $T$. However, this is 
different from the filling, $n=0.88$, at which the density of states displays particle-hole symmetry.  

We find that due to the $\bk$-dependence of the relaxation time and the electron group velocity, the filling at which the 
thermopower crosses zero does not occur at the filling where the density of states shows a particle-hole symmetry at low 
energies. In Fermi liquid theory,~\cite{Stovneng1990} if we assume constant relaxation time and  group velocity, the thermopower is 
proportional to the derivative of the logarithm of the density of states at the Fermi level. This would suggest a thermopower 
which changes sign as the van Hove singularity crosses the Fermi level.  However, in this approach 
$A(\bk,\omega)^2$ in Eq.~(\ref{eq:EffDOS}) is approximated by $\delta(\omega-\epsilon_{\bk})\tau$, where $\tau=\tau_{\bk}$ is 
a $\bk$-independent relaxation time. In addition, the electron group velocity we use in our calculation also has a $\bk$-dependence:
\begin{equation}
v^{x}(\bk)=\frac{\partial \epsilon_{\bk}^{0}}{\partial \bk_x}=2t\sin k_{x}+4t'\sin k_{x}\cos k_{y}.
\label{eq:group_vx}
\end{equation}
If we compare the quantity ${\cal{D}}_{xx}(\omega)$ (the inset of Fig~\ref{fig:Thermopowerfilling}) and the density of states (\Fig{fig:DOS}) 
for different fillings and the same $t'=0$, we find that the effect of $v^{x}(\bk)^2$ is to pull the peak of the density of states to the 
left, because the $\sin k_{x}$ term suppresses the contribution of the van Hove singularity at $X(\pi,0)$ and enhances the contribution 
from the states below the Fermi level. As a result, the thermopower changes sign continuously and at a filling where the quantity 
${\cal D}(\omega)$ has a zero slope at the Fermi level, around $n=0.85$.  As noted by \citebyname{Phillips2010}, this filling is different 
from the one where the density of states displays particle-hole symmetry, $n=0.88$ for $t'=0$.   The impact of the van Hove singularity on 
the thermopower and other transport coefficients is diminished by the fact that the van Hove singularity comes predominantly from a region 
in $\bk$ space where the group velocity goes through zero.

\section{Discussion}  
\label{sec:Discussion}
The results presented here have implications for both the quantum critical phase diagram and the proximity of the superconducting 
dome to the quantum critical point.  

The close proximity of the quantum critical filling and the filling where the van Hove singularity crosses the Fermi level suggests that 
the two are related or even concomitant.  Near the quantum critical point, we find that the flat part of the dispersion orthogonal to the 
antinodal direction is pinned to the Fermi level,  but not the dispersion along that direction (\Fig{DispNearSadPt_blup}).
We also find that the low energy density of states exhibits particle-hole symmetry 
(\Fig{fig:DOS}).   The linear-$T$ resistivity and self energy, characteristic of a marginal Fermi liquid, are observed for the same fillings 
where this pinning is observed. Within the Dynamical Mean Field Approximation (DMFA)~\cite{Majumdar1996}, it is known that if the van Hove 
singularity is pinned to the Fermi level for the non-interacting case, it remains pinned even for the interacting case due to the momentum 
independent nature of the self-energy. In addition, a van Hove singularity initially away from the Fermi level will tend to move towards 
the Fermi level due to the narrowing of the coherent component of the band resulting from electronic correlations.  In the simplest Fermi liquid 
picture, the coherent part of the single-particle Green function $G(\bk,\omega) = Z(\bk)/(\omega+i0^+ - Z(\bk) \epsilon (\bk))$.  So, if 
$Z(\bk)$ becomes small for some values of $\bk$, then we would expect to see a flattening of the observed quasiparticle dispersion, 
$Z(\bk) \epsilon (\bk)$, accompanied by a shift of the peak towards the Fermi level with a vanishing weight.  The new result of 
our work is that the non-local correlations included in the DCA, but absent in the DMFA, are able to move the van Hove singularity, with 
finite weight, to and even across the Fermi level.  This cannot be due solely to the narrowing of the coherent part of the band, since the 
van Hove singularity crosses the Fermi level where quasiparticle fraction $Z$ is already zero.  

Neither a non-interacting picture of the van Hove singularity can completely describe the superconducting transition in
the vicinity of the critical filling.  In a BCS superconductor, the transition is driven by a logarithmic divergence of the bare 
pairing bubble as the temperature falls.  In a recent work,~\cite{Yang2010} the bare $d$-wave pairing susceptibility $\chi'_{0d}$ 
of the 2D Hubbard model was found to diverge algebraically as  $\displaystyle\frac{1}{\sqrt{T}}$ at the quantum critical filling, 
instead of logarithmically, giving rise to a higher \tc.  In the simulation, we traced the origin of this algebraic behavior to 
a component of the dynamic bare bubble which scaled as $\chi''_{0d}(\omega)/\omega = T^{-3/2}  H(\omega/T) $ (see \Fig{chipp_DCA}).  
A van Hove singularity is known to enhance the divergence of the bare pairing bubble.  Since the bare d-wave pairing susceptibility
is dressed only by the self energy, 
with no vertex corrections, a van Hove singularity seems to be the most likely explanation of its enhanced divergence.  However, 
we found that a simple non-interacting picture with a van Hove singularity at the Fermi level does not completely explain the 
observed phenomena.   The standard quadratic dispersion gives a logarithmic divergence of $\chi'_{0d}$ for the half filled model.  
A hypothetical quartic dispersion yields the observed algebraic divergence for $\chi'_{0d}$, but does not give the correct scaling 
for the imaginary part of the bare susceptibility found in \citebyname{Yang2010}. The latter is a consequence of the proximity to 
a quantum critical point, but not necessarily part of the van Hove scenario.

\begin{figure}
\begin{center}$
\begin{array}{cccc}
\includegraphics[width=1.9cm]{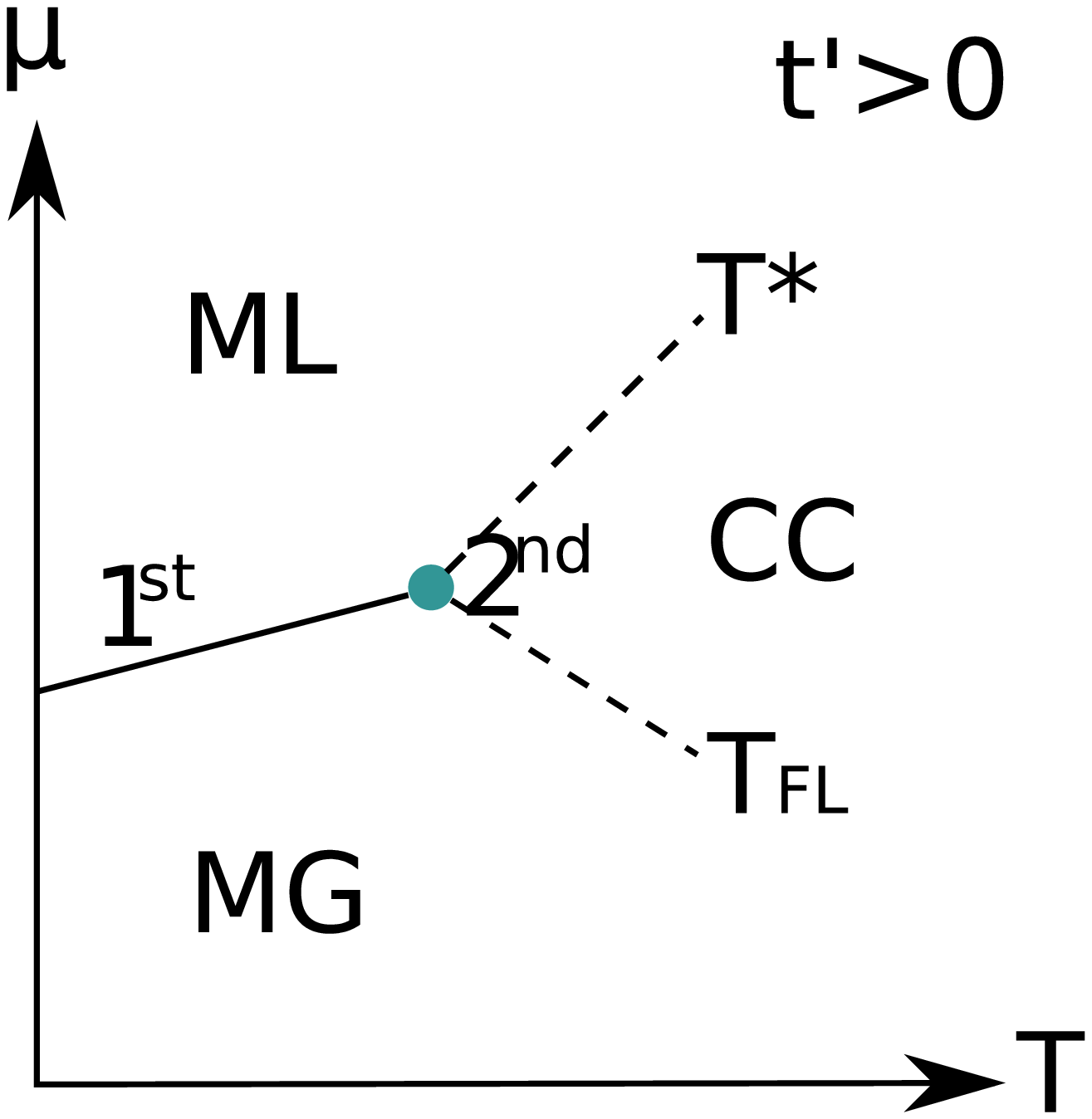} &
\includegraphics[width=1.9cm]{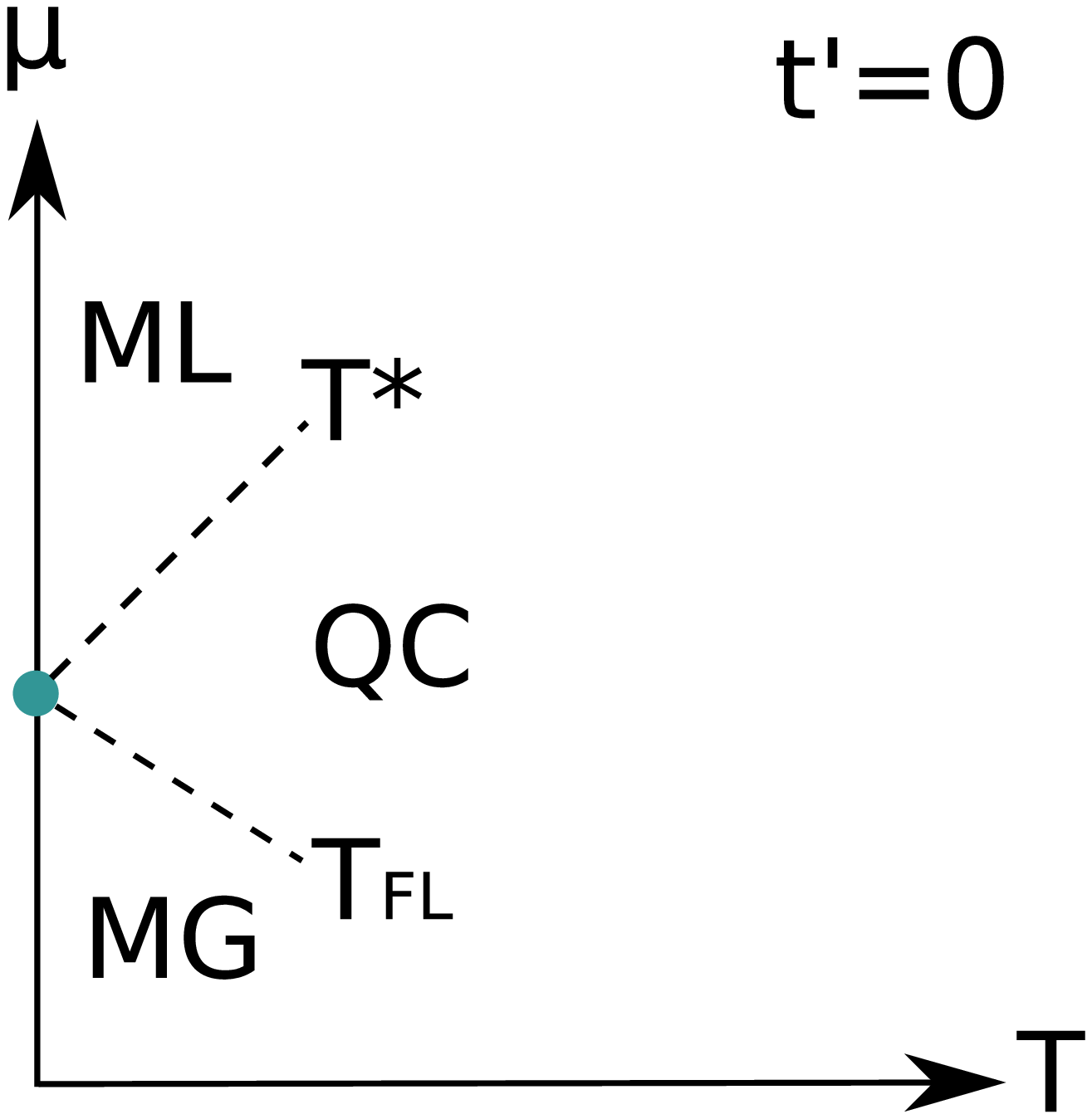} &
\includegraphics[width=1.9cm]{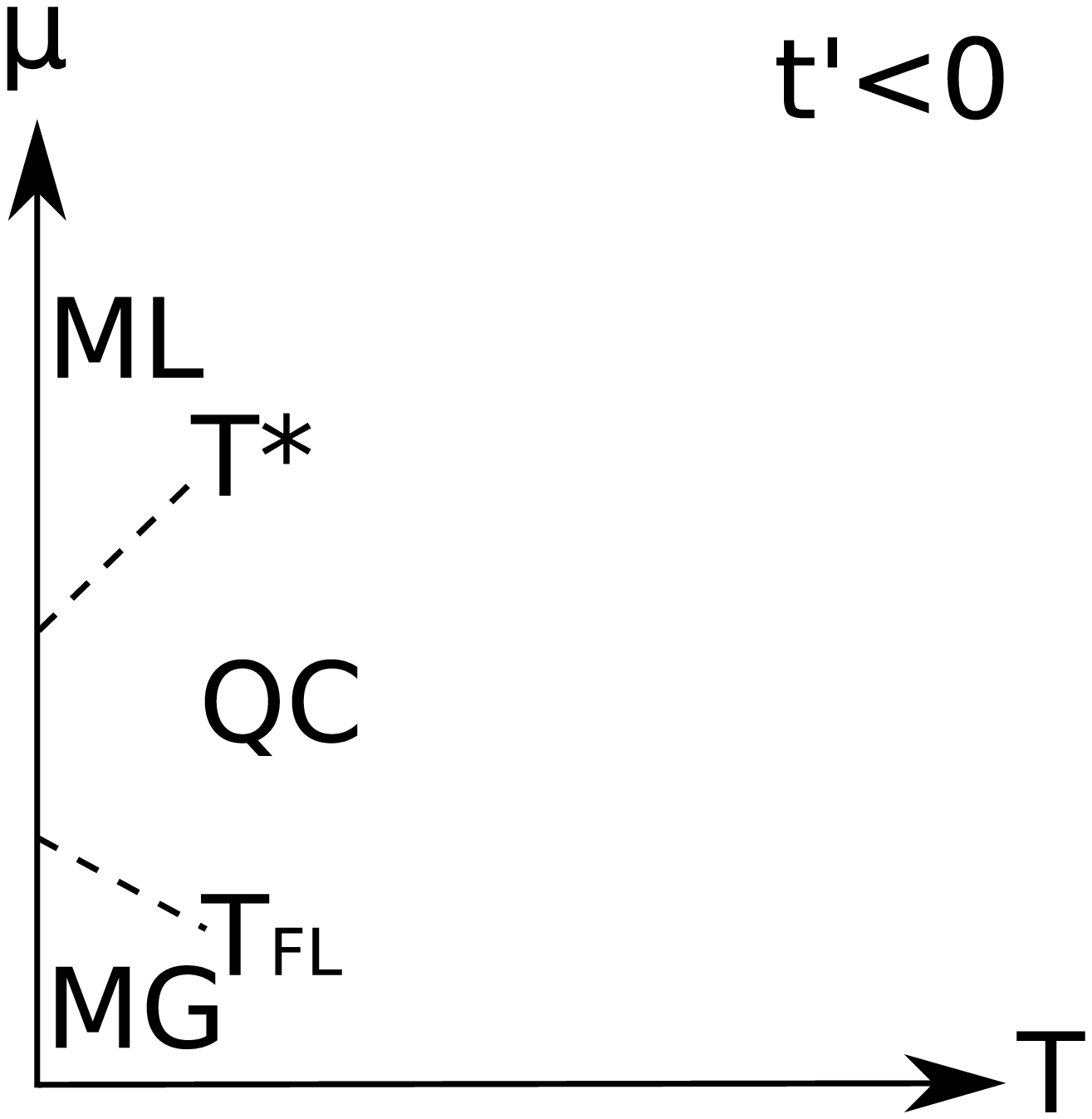} &
\includegraphics[width=2.45cm]{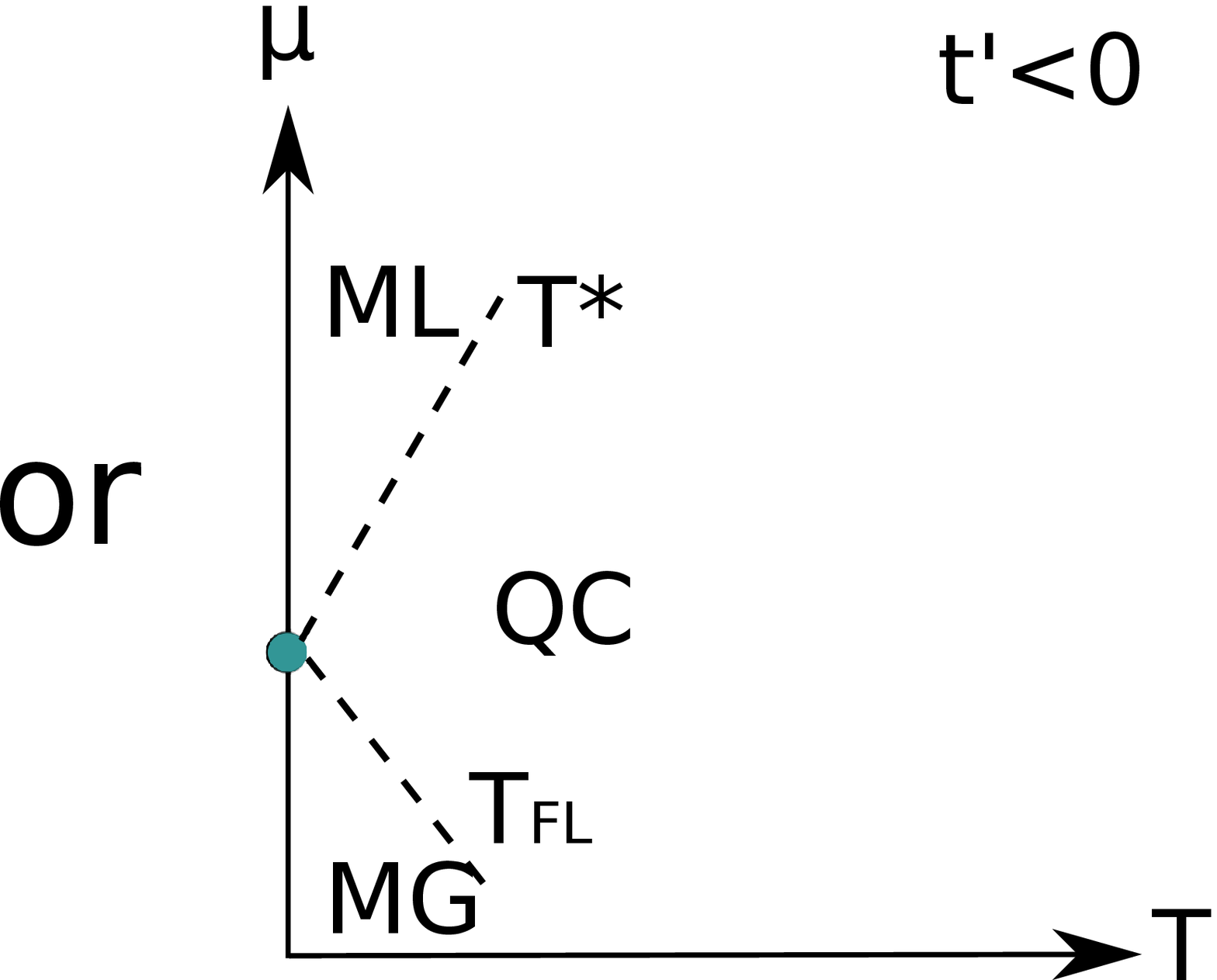}
\end{array}$
\end{center}
\caption{Schematic chemical potential-temperature ($\mu-T$) phase diagram for three values of 
$t'$: (from left to right) $t'>0$, $t'=0$, $t'<0$ scenario presented in this paper and a $t'<0$ scenario based on a Lifshitz transition. CC (QC) indicates the classical 
(quantum) critical region. ML (MG) indicates the Mott liquid (gas) region. $T^*$ is the crossover 
temperature between the Mott liquid and the critical region,  $T_{FL}$ separates the 
Mott gas from the critical region. Note here, we have 
ignored other phases to focus attention on the quantum critical region.}
\label{fig:Phasediagram}
\end{figure}

Our results also shed additional light on the quantum critical phase diagram.  We found previously~\cite{Macridin2006,Khatami2010} 
that when $t'>0$, there is a first order phase separation transition with a first-order line of co-existence in the $\mu-T$ phase 
diagram as shown in Fig.~\ref{fig:Phasediagram}.  In analogy with liquid-gas mixtures, we identify the two phases as a Mott 
liquid, which is insulating and incompressible with well formed local moments and short ranged order, and a Mott gas, which is a weakly
compressible metallic Fermi liquid.  The two phases have the same symmetry, so the first order line of co-existence terminates at 
a second order point where the charge susceptibility diverges.  In analogy to other liquid-gas mixtures, there is a fan-shaped region 
dominated by fluctuations for temperatures above the critical point with neither liquid nor gas character.  When $t'\to 0$, the critical 
point goes continuously to zero temperature and thus becomes a quantum critical point\cite{Khatami2010}.  Above the quantum critical 
point, the marginal Fermi liquid region is found to exist in the V-shape quantum critical region. Inside this region, the only scale  
is the temperature (like Eq.~(\ref{eq:MFL_chi}) and (\ref{eq:MFL_Sigma})). $T^*$  is the temperature separating the marginal Fermi 
liquid from the pseudo-gap region. $T^*$ does not separate the quantum critical region from a region of hidden order.   Rather in 
our scenario, it is only the boundary of the quantum critical region. As we cross from the quantum critical region to the Mott liquid, 
the character of the Mott liquid becomes apparent including a pseudo-gap. 

Here we consider the effect of a negative next-near-neighbor hopping $t'$ on the single particle dispersion.  We find that for 
$t'/t=-0.1$, the flat part of the dispersion orthogonal to the antinode remains in close proximity to the Fermi level for a 
larger range of fillings ($0.83\le n \le 0.87$) when compared with the $t'=0$ result. We find that the resistivity displays 
a linear-$T$ dependence, and the self energy displays MFL characteristics in the same range of doping.  We also note
that the single-particle spectra is less coherent in the center of this doping region than it is at the quantum 
critical filling when $t'=0$.  These observations are consistent with an increase in the doping region of marginal
Fermi liquid character when $t'/t < 0$.  

There are several different scenarios which may explain this behavior.  Since the doping region where the van Hove singularity 
is near the Fermi level increases with decreasing $t'/t$, perhaps the simplest understanding of this behavior is that the van 
Hove singularity \emph{pinning} gives rise to the marginal Fermi liquid behavior\cite{Pattnaik1992, Newns1994, Tsuei1990}.  Another 
possibility is that both $T$ and negative $t'$ are relevant variables, or variables that, when finite, move the system away from a 
critical point.  In addition, they are roughly similar in their effect, in that when $t'=T=0$ the doping region of marginal Fermi 
liquid character shrinks to a point, but the region increases with either increasing $T$ or decreasing $t'$.  Thus when $t'<0$, the 
quantum critical point may be viewed as moving to negative temperatures so that the quantum critical region broadens to allow the 
linear-$T$ resistivity and the pinning of the flat $k_y$ dispersion to the Fermi level to exist for a wider range of fillings 
(compare e.g. the third panel of \Fig{fig:Phasediagram} and inset of \Fig{chipp_DCA}).  However, this interpretation is not consistent 
with scaling theory where we expect a finite low temperature scale like $T_F$ or $T^*$ to emerge for any set of parameters that 
take us away from the quantum critical point.  Another possibility is that the change in Fermi surface topology associated with 
the van Hove singularity crossing the Fermi level can introduce \emph{Lifshitz} singularities in the free 
energy\cite{Imada2005,Takahiro2006,Yamaji2006,Takahiro2007,Yamaji2007,Imada2010,Norman2010}. 
This scenario will mean a line of zero temperature critical points in the $t'-\mu$ plane beyond the quantum critical point as the 
control parameter $t'$ is decreased from zero to negative values in the $t'-\mu$ plane.  For positive $t'$ the Lifshitz
transition may yield a line of first order critical points in $t'-\mu$ plane which terminates at the quantum critical point for $t'=0$.  
To understand our results, the line of Lifshitz transitions for negative values of $t'/t$ must yield an associated V-shaped region 
of quantum criticality which becomes flatter as $t'$ decreases as shown on the right in Fig.~\ref{fig:Phasediagram}.  
Finally, another mechanism enhanced by the van Hove singularity is the \emph{Pomeranchuk instability}\cite{Halboth2000a,Yamase2005} of 
the Fermi surface where the Fermi surface is distorted to break the $C_4$ symmetry of the square lattice. The possibility of 
Lifshitz and Pomeranchuk transitions are being studied currently and will be published elsewhere.  

There is some experimental evidence\cite{Kaminski2006} in the cuprates that there is a change in Fermi surface topology and an 
associated Fermi level crossing of the van Hove singularity at a doping that is larger than the doping at which \tc is
maximum, while still being within the dome. On the other hand, we find that the van Hove singularity crosses the Fermi level 
at a slightly smaller doping than the optimal doping. This disagreement can be due to the other effects present in the real 
systems e.g. phonons that are not incorporated in this model calculation, or the strong doping dependence of the strength 
of the pairing interaction\cite{Yang2010} seen in the simulations. 

The transport provides some additional evidence for the van Hove singularity.  In our calculations, the low energy particle-hole 
symmetry and the change in sign of the thermopower with doping near the critical value are both due to the crossing of the 
van Hove singularity.  However, the doping associated with the van Hove crossing differs from that where the thermopower is 
zero due to the anisotropy of the group velocity on the Fermi surface.

\section{Conclusion}  
\label{sec:Conclusion}
We explore the role of the van Hove singularity in the quantum criticality observed at finite doping in the 
Hubbard model.\cite{Vidhyadhiraja2009}  Near the quantum critical filling, we find a van Hove singularity 
due to a flattening of the dispersion near the Fermi level.  The motion of the flat part of the dispersion 
along the antinodal direction is anisotropic.  The part along the antinode moves continuously through the 
Fermi level.  The part orthogonal to this direction is pinned near the Fermi level at a filling where the 
self energy, transport, and energies\cite{k_mikelsons_09b} also display marginal Fermi liquid behavior, 
and the quasiparticle fraction vanishes.\cite{Vidhyadhiraja2009}.  Many authors have proposed that the van 
Hove singularity near the Fermi level will enhance superconductivity by enhancing the divergence of the bare pairing 
bubble.  Indeed we found previously that the superconducting dome surrounds the critical doping where the real part of 
the pairing bubble diverges algebraically, replacing the Fermi liquid log divergence.\cite{Yang2010}  However, a simple 
non-interacting picture with the van Hove singularity at the Fermi level doesn't explain the quantum critical 
scaling of the bare dynamic pairing susceptibility.  We also found previously that a positive $t'$ is the 
control parameter for a first order phase separation transition which is terminated by a second order critical
point. As $t'\to 0$ this second order terminus is driven to zero temperature yielding the quantum critical
point.\cite{Khatami2010}  Here we explore the effect of a negative $t'$, and find that it is a relevant
variable which increases the extent in doping (and chemical potential) of the quantum critical region characterized 
by marginal Fermi liquid behavior.

\begin{acknowledgments}
We would like to thank Piers Coleman, Jan Zaanen, Thomas Pruschke, H. R. Krishnamurthy, Sebastian Schmitt, George Kastrinakis, 
Ka-Ming Tam, Matthias Vojta, Masa Imada, Walter Metzner and Jian-Huang She for useful conversations. We also thank Robert Markiewicz 
for careful reading of the manuscript and providing insightful suggestions. This work is supported by DOE SciDAC DE-FC02-06ER25792 
(SP, SXY, KM, MJ),  NSF grants OISE-0952300 (KSC, SXY, SQS, JM) and DMR-0706379 (DG and MJ).  Supercomputer support was provided 
by the NSF  TeraGrid under grant number TG-DMR100007.  This research also used resources of the National Center for Computational 
Sciences at Oak Ridge National Laboratory, which is supported by the Office of Science of the U.S. Department of Energy under 
Contract No.\ DE-AC05-00OR22725.
\end{acknowledgments}


\bibliography{BIB_SP}

\end{document}